\documentclass[prb,twocolumn,superscriptaddress,amsmath,amssymb]{revtex4-2}
\usepackage[colorlinks=true,citecolor=blue,linkcolor=blue,breaklinks=true]{hyperref}

\usepackage{color,graphicx}
\usepackage{bm}
\usepackage{amsmath}
\usepackage{amssymb}
\usepackage{caption}
\usepackage[usenames,dvipsnames]{xcolor}
\usepackage{verbatim}
\usepackage{float}
\usepackage[justification=Justified]{caption}

\begin{document}
	
	\title{Germanium-based hybrid semiconductor-superconductor topological quantum computing platforms: Disorder effects}
	
	\author{Katharina Laubscher}
	
	\author{Jay D. Sau}
	
	\author{Sankar Das Sarma}
	
	\affiliation{Condensed Matter Theory Center and Joint Quantum Institute, Department of Physics, University of Maryland, College Park, MD 20742, USA}
	
	\date{\today}
	
	\begin{abstract}
It was recently suggested that proximitized gate-defined Ge hole nanowires could serve as an alternative materials platform for the realization of topological Majorana zero modes (MZMs). Here, we theoretically study the expected experimental signatures of Ge-based MZMs in tunneling conductance measurements, taking into account that unintentional random disorder is unavoidably present in realistic devices. Explicitly, we present numerically calculated local and nonlocal tunneling conductance spectra (as functions of bias voltage and magnetic field) for two different wire lengths, two different disorder models, two different parent superconductors (Al and NbTiN), and various disorder strengths, which we relate to an estimated lower bound for the disorder strength in Ge-based hybrid devices that we extract from the experimentally reported hole mobilities in current state-of-the-art two-dimensional Ge hole gases. We find that even if the actual disorder strength in the Ge-based hybrid device exceeds this theoretical lower bound by an order of magnitude, the system is still in the weak disorder regime, where the topological superconductivity is intact and the local conductance spectra manifest clear end-to-end correlated zero-bias peaks if and only if the system hosts topological end MZMs. This shows that, despite the relatively small pristine topological gaps ($\sim\,$a few tens of $\mu$eV), Ge-based hybrid devices are a promising alternative platform for Majorana experiments due to the extremely high materials quality, which leads to an increased gap-to-disorder ratio and, therefore, to less ambiguity in the experimental tunneling conductance data compared to InAs-based devices.
	\end{abstract}
	
	\maketitle

	\section{Introduction}

It is by now well-known that disorder is one of the main obstacles preventing the realization and conclusive observation of topological Majorana zero modes (MZMs) in superconductor-semiconductor (SC-SM) hybrid devices~\cite{Dassarma2023}. Soon after the initial theory works predicted that topological MZMs can be realized in semiconductor nanowires with strong spin-orbit interaction (made from, e.g., InAs or InSb) contacted by an ordinary superconductor (such as, e.g., Al or Nb)~\cite{Lutchyn2010,Stanescu2011,Oreg2010}, several experimental groups reported the observation of the key MZM signature in these devices, namely, a zero-bias peak (ZBP) in the local tunneling conductance ~\cite{Mourik2012,Rohkinson2012,Das2012,Deng2012,Lee2012,Churchill2013,Deng2016,Deng2018}. While this was interpreted as compelling evidence for the presence of topological end MZMs at the time, it was subsequently realized that the situation is significantly complicated by the presence of substantial disorder in the experimental samples. Indeed, if strong enough, disorder both destroys the topological MZMs and, at the same time, induces trivial fermionic subgap states (Andreev bound states) that can accidentally mimick the signatures of MZMs (including, most importantly, ZBPs~\cite{Sau2012,Sau2013,Chiu2017,Pan2020,Pan2020b,Pan2021,Dassarma2021,Ahn2021,Dassarma2023}). Because of this ambiguity in the experimental tunneling conductance data, conclusive evidence for the observation of MZMs has not yet been obtained up to date, and the consensus in the field seems to be that most (if not all) ZBPs that have been observed so far are of trivial origin.

Intense experimental efforts, both in physics and in materials sciences, have been made (and are still being made) to improve the quality of the InAs platform, as it seems likely that MZMs cannot be observed unless the amount of disorder is significantly reduced. Substantial progress in this direction has recently been made in an experiment by Microsoft~\cite{Aghaee2022}, which reported the observation of small and fragile topological gaps in a new generation of gate-defined InAs/Al hybrid nanowires. A detailed theoretical analysis~\cite{Dassarma2023b,Dassarma2023c} shows that the disorder levels in these state-of-the-art InAs-based devices are likely one order of magnitude lower than the disorder levels in previous Majorana experiments, but still not low enough to allow for an unambiguous identification of MZMs based on tunneling spectroscopy data. As such, significant further improvement of the materials quality is crucial if topological MZMs should be realized in InAs-based devices.

A more radical approach would be to move away from InAs altogether, replacing it by an alternative semiconductor with a higher intrinsic materials quality. In this context, two-dimensional hole gases (2DHGs) in Ge~\cite{Scappucci2021} could be a promising candidate platform. Indeed, Ge 2DHGs can be fabricated in ultra-high quality with mean-free paths on the order of tens of microns and peak mobilities on the order of millions~\cite{Sammak2019,Lodari2022,Myronov2023,Stehouwer2023}. Furthermore, Ge holes exhibit a sizable spin-orbit interaction, and proximity-induced superconducting gaps (some of them hard) have recently been observed in Ge/SC hybrids with Al, Nb, or PtSiGe as the parent superconductor~\cite{Vries2018,Hendrickx2018,Hendrickx2019,Vigneau2019,Aggarwal2021,Tosato2022,Valentini2024}. (An important open question is of course whether the high mobilities of the bare Ge quantum wells can be maintained in proximitized hybrid devices, but this is a question that ultimately only materials scientists can answer, transcending the theoretical issues considered in the current work.) A recent theoretical study~\cite{Laubscher2023} shows that the realization of topological MZMs in gate-defined Ge hole nanowires is indeed a feasible idea if sufficiently narrow and clean hybrid nanowires can be fabricated, the main disadvantage however being the small $g$ factor, which limits the maximal topological gaps to a few tens of $\mu$eV in a pristine system. Nevertheless, since the important dimensionless quantity describing the quality of SC-SM hybrid Majorana nanowires is the gap-to-disorder ratio (i.e., the ratio between the topological gap and the disorder strength), the Ge-based platform may still bring substantial advantages over the InAs-based platform due to the extremely high materials quality of the already existing Ge samples. Furthermore, the problems caused by the small $g$ factor can potentially be mitigated if a parent superconductor with a large critical magnetic field is used (such as, e.g., Nb).

In the present work, we analyze the situation in the Ge platform in more detail by presenting numerical results for the local and nonlocal tunneling conductance spectra in gate-defined Ge hole nanowires in the presence of unintentional random disorder in the chemical potential. We present results for two different wire lengths, two different disorder models, two different parent superconductors (Al and NbTiN), and various disorder strengths, based on which we critically discuss the potential advantages of the Ge-based platform as compared to the InAs platform. Based on experimentally reported peak mobilities, we estimate the disorder strength in current state-of-the-art Ge 2DHGs to be on the order of a couple of $\mu$eV, which provides a natural lower bound for the disorder strength in Ge-based hybrid devices, where additional disorder is introduced by additional processing steps and the existence of interfaces. Our numerical results show that even for disorder strengths that exceed this theoretical lower bound by an order of magnitude, the local conductance spectra remain almost indistinguishable from the pristine case, manifesting clear MZM-induced ZBPs when the system is in the topological phase and, perhaps even more importantly, no trivial disorder-induced ZBPs in the trivial phase. Therefore, our work suggests that the weak disorder regime suitable for the realization and observation of topological MZMs may well be accessible in Ge hole nanowires fabricated from state-of-the-art Ge 2DHGs. This should be compared with the current situation in InAs/Al hybrid devices~\cite{Aghaee2022}, where the disorder strength in the parent two-dimensional electron gas (2DEG) can be estimated to be $\sim 0.5-1$~meV, which is already by itself at least $3$ times larger than the maximal pristine topological gap. Consistent with this estimate, detailed numerical simulations~\cite{Dassarma2023c,Dassarma2023b} have shown that even the best InAs-based SC-SM hybrid nanowires are likely still in the intermediate disorder regime.
	
This paper is organized as follows. In Sec.~\ref{sec:model}, we describe our model of the Ge/SC hybrid device. In particular, we describe the Ge hole nanowire in Sec.~\ref{sec:model_ge}, the disorder potential in Sec.~\ref{sec:model_disorder}, and the three-terminal device used to simulate local and nonlocal conductance measurements in Sec.~\ref{sec:model_three_terminal}. In Sec.~\ref{sec:conductance}, we then present our numerical results for the local and nonlocal tunneling conductance in Ge/SC hybrid nanowires as a function of bias voltage $V_\mathrm{bias}$ and magnetic field $B$ in the presence of disorder. We present results for two different wire lengths, two different disorder models (random Gaussian disorder and spatially correlated disorder caused by random charged impurities), two different parent superconductors (Al and NbTiN), and various disorder strengths. We critically discuss our numerical results in the context of an estimated lower bound for the disorder strength in state-of-the-art Ge 2DHGs, showing that the Ge-based Majorana platform has substantial potential advantages over the InAs-based platform if sufficiently clean and narrow Ge hole channels can be fabricated from the already existing high-quality Ge 2DHGs. Finally, we conclude in Sec.~\ref{sec:conclusions}.
	
	 \begin{figure*}[bt]
		\centering
		\includegraphics[width=\textwidth]{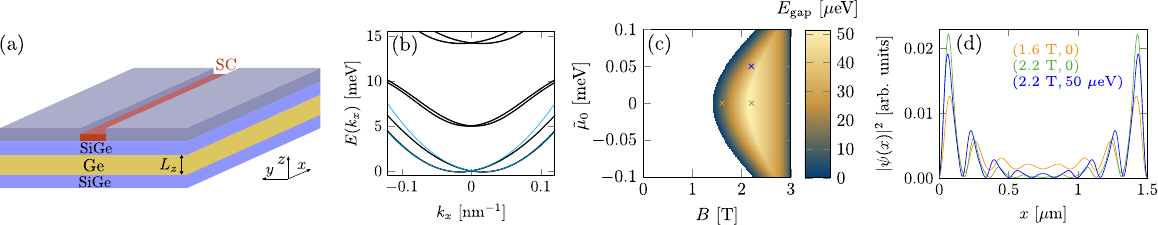}
		\caption{(a) We consider a quasi-1D Ge hole nanowire defined by electrostatic gates (gray) placed on a Ge quantum well of thickness $L_z$ (yellow) sandwiched between two layers of SiGe (blue). If the nanowire is additionally proximitized by a superconductor (red) and a magnetic field is applied along the wire axis, MZMs can emerge at the wire ends. (b) Low-energy spectrum of a bare Ge hole nanowire at $B=0$ (note that a global minus sign is omitted from the hole spectrum throughout this work). The spectrum of the full (effective) model is shown in black (blue). (c) Bulk topological phase diagram for a Ge/Al hybrid nanowire. The white (colored) regions correspond to the trivial (topological) phase. We note that this phase diagram was obtained without taking the self-energy effect of the parent superconductor into account. If the self-energy effect is taken into account, the topological gaps are reduced by a factor of $2-3$ compared to what is shown here and the topological phase transition is shifted to larger magnetic fields (see below). (d) Majorana wave functions in a finite wire of length $L_x=1.5~\mu$m for specific values of magnetic field and chemical potential indicated by the crosses in (c) and listed in the inset. Again, these wave functions were calculated without taking the self-energy effect into account. In (b-d), the parameters for the Ge hole nanowire are $L_y=15$~nm, $L_z=22$~nm, $\mathcal{E}=0.5$~V$/\mu$m, and $E_s=0.01$~eV. The chemical potential $\tilde\mu_0$ is measured from the spin-orbit point of the lowest confinement-induced subband. In (c) and (d), the proximity-induced superconducting gap at zero magnetic field is set to $\Delta(0)=80~\mu$eV, and the critical field of the Al strip is taken to be $B_c=3$~T. }
		\label{fig:setup}
	\end{figure*}

	\section{Model}
	\label{sec:model}

	\subsection{Ge hole nanowire}
	\label{sec:model_ge}
	
	\subsubsection{Multi-band model}
	\label{sec:model_ge_full}
	
	We consider a gate-defined Ge hole nanowire in a Ge/SiGe quantum well proximitized by a superconductor (such as, e.g., Al), see Fig.~\ref{fig:setup}(a) for a schematic. A previous theoretical study~\cite{Laubscher2023} has shown that such a hybrid Ge hole nanowire can enter a topological superconducting phase with MZMs at the wire ends. Since the model used to describe the pristine Ge hole nanowire has already been discussed in detail in Ref.~\cite{Laubscher2023}, we will only give a brief summary here and refer the interested reader to Ref.~\cite{Laubscher2023} for additional explanations. We start by describing the bare Ge hole nanowire, which we model by a Hamiltonian $H_0=\int d\bm{r}\,\psi^\dagger(\bm{r})\mathcal{H}_0(\bm{r})\psi(\bm{r})$ with $\psi=(\psi_{3/2}$,$\psi_{1/2},\psi_{-1/2},\psi_{-3/2})^T$ and
	\begin{align}
	\mathcal{H}_0&=\frac{\hbar^2}{m}\left[
	\left(\gamma_1 + \frac{5 \gamma_s}{2}\right)\frac{\bm{k}^{2}}{2}
	- \gamma_s \left( \bm{k} \cdot \bm{J} \right)^2 
	\right]-\mu_0\nonumber\\&\quad+V(y,z)-E_s J_z^2-e \mathcal{E} z+H_\mathbf{B}.\label{eq:normalH}
	\end{align}
	The first line of Eq.~(\ref{eq:normalH}) is the isotropic Luttinger-Kohn (LK) Hamiltonian, which describes the topmost valence band holes of 3D bulk Ge~\cite{Luttinger1956,Winkler2003}. Here, $m$ denotes the bare electron mass, $\gamma_1=13.35$, $\gamma_2=4.25$, and $\gamma_3=5.69$ are the Luttinger parameters for Ge, $\gamma_s=(\gamma_2+\gamma_3)/2$, $\bm{k}=(k_x,k_y,k_z)$ is the vector of momentum, $\bm{J}=(J_x,J_y,J_z)$ is the vector of spin-$3/2$ operators, and $\mu_0$ is the chemical potential. In the second line of Eq.~(\ref{eq:normalH}), the confinement potential $V(y,z)$ describes (i) the confinement along the $z$ direction arising due to the SiGe barrier defining the quantum well and (ii) the gate-induced confinement along the $y$ direction that defines the quasi-1D nanowire. We model this confinement potential as an infinite square well in both the $y$ and $z$ direction,
	\begin{equation}
	V(y,z)=\begin{cases} 0 & |y|<L_y/2, \ |z|<L_z/2,\\\infty & \mathrm{otherwise,}\end{cases}\label{eq:conf}
	\end{equation}
	where $L_y$ is the width of the 1D channel and $L_z$ is the thickness of the quantum well. It has previously been shown that using, e.g., a parabolic potential to model the gate-induced confinement along the $y$ direction leads only to quantitative, but not qualitative, changes in the topological phase diagram of the Ge hole nanowire~\cite{Laubscher2023}, which is why we focus on the simplest (and most optimistic) scenario of infinite hard-wall confinement here.
	Furthermore, $E_s>0$ is the strain energy arising due to the lattice mismatch between the Ge and the SiGe barrier~\cite{Bir1974}, and $\mathcal{E}$ is an electric field applied along the $z$ direction (i.e., the out-of-plane direction), which gives rise to spin-orbit interaction. Additionally, $H_\mathbf{B}=\mathcal{H}_Z+\mathcal{H}_\mathrm{orb}$ describes the effect of a magnetic field of strength $B$ applied along the $x$ direction, i.e., along the nanowire axis. Here, the Zeeman term $\mathcal{H}_Z$ takes the form $\mathcal{H}_Z=2\kappa\mu_B B J_x$~\cite{Luttinger1956,Winkler2003}
	with $\mu_B$ the Bohr magneton and $\kappa\approx 3.41$ for Ge~\cite{Lawaetz1971}. The orbital effects associated with the magnetic field lead to an extra term $\mathcal{H}_\mathrm{orb}$ in the bulk LK Hamiltonian~\cite{Adelsberger2022}
	\begin{align}
	&\mathcal{H}_\mathrm{orb}=\frac{\hbar e}{2m}\left[\frac{2\gamma_1+5\gamma_s}{2}\left(\frac{e}{\hbar}\bm{A}^2+2\bm{k}\cdot\bm{A}\right)-\frac{2\gamma_s e}{\hbar}\left(\bm{A}\cdot\bm{J}\right)^2\right.\nonumber\\& -4\gamma_s\big(k_xA_xJ_x^2+\left(\{k_x,A_y\}+\{k_y,A_x\}\right)\{J_x,J_y\}+\mathrm{c.p.}\big)\bigg],\label{eq:orb}
	\end{align}
	where $\bm{A}$ is the vector potential satisfying $\bm{B}=\bm{\nabla}\times\bm{A}$, $\{A,B\}=(AB+BA)/2$, and where `c.p.' stands for `cyclic permutations'. For our numerical simulations, we fix the gauge to $\bm{A}=(0,0,By)$. 
	
	Finally, we incorporate a proximity-induced superconducting pairing term into our model. Since the precise microscopic description of the proximity-induced superconducting pairing in Ge/SC hybrid structures is not known and, in addition, is likely to depend on details of the specific setup, we work with a very simple pairing term of the form
	\begin{equation}
	H_{sc}=\int d\bm{r} \sum_{s=\frac{1}{2},\frac{3}{2}}\Delta_s\,\psi_s^\dagger(\bm{r}) \psi_{-s}^\dagger(\bm{r})+\mathrm{H.c.},\label{eq:sc_pairing}
	\end{equation}
	where $\Delta_{1/2}$ is the superconducting pairing amplitude for holes with spin projection $\pm 1/2$ along the $z$ direction (light holes, LHs) and $\Delta_{3/2}$ is the pairing amplitude for holes with spin projection $\pm 3/2$ along the $z$ direction, (heavy holes, HHs). To keep the number of unknown parameters to a minimum, we further assume that the LH and HH pairing amplitudes are equal in magnitude but of opposite sign, i.e., $\Delta_{3/2}=-\Delta_{1/2}\equiv\Delta$. However, it is straightforward to extend our analysis to different pairing amplitudes for HHs and LHs. Within our model, this would simply lead to a geometry-dependent renormalization of the effective superconducting gap that is opened in the lowest confinement-induced hole subband (see also below).
	
	Additionally, we phenomenologically model the suppression of the proximity-induced superconducting gap due to the applied magnetic field as
	\begin{equation}
	\Delta(B)=\Delta(0)\sqrt{1-\left(\frac{B}{B_c}\right)^2}\Theta(B_c-|B|),\label{eq:scgap_B}
	\end{equation}
	where $\Delta(0)$ is the proximity-induced superconducting gap at zero magnetic field, $B_c$ is the critical magnetic field of the superconductor, and $\Theta$ is the Heaviside step function. The total multi-band Hamiltonian describing the proximitized Ge hole nanowire is then given by $H=H_0+H_{sc}$. This Hamiltonian can be solved numerically by projecting onto a suitable set of low-energy eigenfunctions of the infinite square well potential given in Eq.~(\ref{eq:conf}), which is a technique that is frequently used in the literature on hole systems~\cite{Csontos2009,Kloeffel2011,Kloeffel2018,Adelsberger2021,Adelsberger2022,Milivojevic2021} and therefore only briefly summarized in Appendix~\ref{app:parameters}.

	\subsubsection{Effective low-energy description}
	\label{sec:model_ge_eff}
	
	Throughout this work, we focus on the regime where only the lowest confinement-induced subband of the normal-state Hamiltonian $H_0$ [see Eq.~(\ref{eq:normalH})] is occupied. In this regime, the band structure of $H_0$ can be described by an effective low-energy model that only includes two bands (labeled by a pseudospin index $\sigma\in\{\uparrow,\downarrow\}$). Indeed, up to a global minus sign that we omit in this work, the low-energy spectrum of $H_0$ [see Fig.~\ref{fig:setup}(b) for an example] resembles the spectrum of electrons in a conventional Rashba nanowire and can be described by a simple effective Hamiltonian of the form $H_0^\mathrm{eff}=\int dx\,\psi^\dagger(x)\mathcal{H}_0^\mathrm{eff}\psi(x)$ with $\psi=(\psi_\uparrow,\psi_\downarrow)^T$ and~\cite{Adelsberger2022}
	\begin{equation}
	\mathcal{H}_0^\mathrm{eff}=-\frac{\hbar^2 \partial_x^2}{2\bar{m}}-\mu+\left(V_Z-\frac{\hbar^2\partial_x^2}{2\bar{m}_s}\right)\sigma_x-i\alpha_{so}\partial_x\sigma_y,\label{eq:eff2band}
	\end{equation}
	where $\bar{m}$ is the effective mass, $\alpha_{so}$ is the effective spin-orbit coupling strength, $\bar{m}_s$ is an effective spin-dependent mass, and the Pauli matrices $\sigma_i$ with $i\in\{x,y,z\}$ act in pseudospin space. Furthermore, we have defined the effective Zeeman splitting $V_Z=g_\mathrm{eff}\mu_BB/2$, where $g_\mathrm{eff}$ is the effective $g$ factor. However, in contrast to spin-$1/2$ electrons in conventional Rashba nanowires, the low-energy holes considered here are a mix of HHs and LHs, with the degree of HH-LH mixing depending sensitively on the details of the system (wire geometry, strain, electromagnetic fields,...). As a consequence, the effective parameters entering Eq.~(\ref{eq:eff2band}) also depend on all of these details~\cite{Adelsberger2021,Adelsberger2022}, such that, for each given set of system parameters, the effective parameters should be recalculated from the full Hamiltonian $H_0$ given in Eq.~(\ref{eq:normalH}). 
		
	In the presence of a superconducting pairing term, the effective Hamiltonian can be written in Bogoliubov-de Gennes (BdG) form as $H^\mathrm{eff}=\frac{1}{2}\int dx\,\psi^\dagger(x)\mathcal{H}^\mathrm{eff}\psi(x)$ with $\psi=(\psi_\uparrow,\psi_\downarrow,\psi_\downarrow^\dagger,-\psi_\uparrow^\dagger)^T$ and
	\begin{align}
	\mathcal{H}^\mathrm{eff}&=\left(-\frac{\hbar^2 \partial_x^2}{2\bar{m}}-\mu\right)\tau_z+\left(V_Z-\frac{\hbar^2\partial_x^2}{2\bar{m}_s}\right)\sigma_x\nonumber\\&\quad\ -i\alpha_{so}\partial_x\sigma_y\tau_z+\Delta(B)\tau_x,\label{eq:2band_sc}
	\end{align}
	where we have introduced Pauli matrices $\tau_i$ for $i\in\{x,y,z\}$ acting in particle-hole space and where $\Delta(B)$ is given in Eq.~(\ref{eq:scgap_B}). Indeed, for our specific choice of pairing amplitudes $\Delta_{3/2}=-\Delta_{1/2}\equiv\Delta$ in Eq.~(\ref{eq:sc_pairing}), the superconducting gap that enters the effective model is just $\Delta$. For a more general choice of pairing amplitudes, the gap in the effective model should be replaced by an effective gap $\Delta_\mathrm{eff}$, the exact value of which then also depends on details such as, e.g., the wire geometry~\cite{Mao2012}. In the following, we will just treat $\Delta$ as a  phenomenological input parameter that conveniently allows us to model the experimentally observed proximity-induced gaps in Ge/SC hybrids, but we do not claim that our description of the superconducting pairing in Eq.~(\ref{eq:sc_pairing}) is microscopically accurate (it is certainly oversimplified).

	Furthermore, the simple description of the proximity-induced pairing used in Eq.~(\ref{eq:2band_sc}) is only valid in the limit of weak SC-SM coupling between the parent superconductor and the Ge 2DHG. One step towards a more realistic description of the  intermediate to strong coupling regime can be made by replacing the pairing term in Eq.~(\ref{eq:2band_sc}) by a self-energy term~\cite{Sau2010,Stanescu2010,Stanescu2011,Cole2015,Reeg2018}
	\begin{equation}
	\Sigma(\omega)=-\gamma\frac{\omega+\Delta_0(B)\tau_x}{\sqrt{\Delta_0^2(B)-\omega^2}},\label{eq:sc_pairing_SE}
	\end{equation}
	where $\omega$ is the energy, $\Delta_0(B)$ is the gap of the parent superconductor, which we again take to be a function of magnetic field following Eq.~(\ref{eq:scgap_B}), and $\gamma$ is the SC-SM coupling strength. The total effective BdG Hamiltonian then becomes energy-dependent and reads
	 \begin{align}
	  \mathcal{H}^\mathrm{eff}_\mathrm{SE}(\omega)&=\left(-\frac{\hbar^2 \partial_x^2}{2\bar{m}}-\mu\right)\tau_z+\left(V_Z-\frac{\hbar^2\partial_x^2}{2\bar{m}_s}\right)\sigma_x\nonumber\\&\quad\ -i\alpha_{so}\partial_x\sigma_y\tau_z+\Sigma(\omega).\label{eq:2band_self}
	 \end{align}
     The SC-SM coupling strength $\gamma$ depends on the LH and HH tunneling amplitudes across the SC-SM interface~\cite{Stanescu2013,Adelsberger2023}, which in turn depend on the exact geometry of the hybrid system and on microscopic details of the interface. (Furthermore, in addition to the direct coupling between the Ge hole bands and the superconductor, there could also be an indirect contribution to the proximity effect mediated by the conduction band~\cite{Moghaddam2014}.) We will make no effort here to understand the tunneling between the Ge 2DHG and the superconductor at a microscopic level as this would be beyond of the scope of this work. Instead, we just take $\gamma$ as an input parameter, the value of which we choose such that the size of the resulting proximity-induced gap is consistent with  experimentally reported values in Ge/SC hybrids~\cite{Vries2018,Hendrickx2018,Hendrickx2019,Valentini2024,Vigneau2019,Aggarwal2021,Tosato2022}.

	 \begin{figure*}[bt]
	 	\centering
	 	\includegraphics[width=\textwidth]{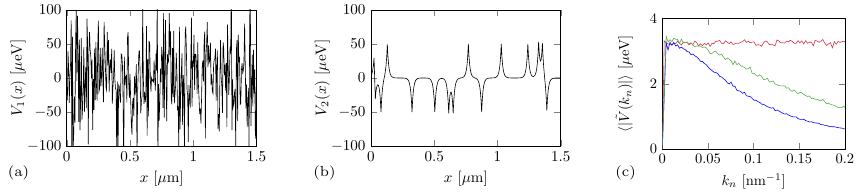}
	 	\caption{(a) Example realization of a random onsite disorder potential drawn from an uncorrelated Gaussian distribution as defined in Eq.~(\ref{eq:disorder_uncorr}) for a wire of length $L_x=1.5~\mu$m. Here, we have used $\sigma=50~\mu$eV and the lattice constant was set to $a=5$~nm. (b) Example realization of a correlated disorder potential as defined in Eq.~(\ref{eq:disorder_corr}) for a wire of length $L_x=1.5~\mu$m. Here, we have used $V_0=50~\mu$eV, $\lambda=11$~nm, and $n_i=10/\mu$m. (c) Spectral signatures $\langle|\tilde{V}(k_n)|\rangle$ corresponding to the random Gaussian disorder model defined in Eq.~(\ref{eq:disorder_uncorr}) (red: $\sigma=50~\mu$eV) and to the correlated disorder model defined in Eq.~(\ref{eq:disorder_corr}) (blue: $\lambda=11$~nm and $V_0=50~\mu$eV, green: $\lambda=7$~nm and $V_0=80~\mu$eV). The other parameters are the same as in (a) and (b), respectively, and we have averaged over $2000$ disorder configurations. The spectral signatures of all three disorder types agree at small wave vectors. The random Gaussian disorder potential is a better (worse) approximation for the correlated disorder potential if $\lambda$ is smaller (larger), see Ref.~\cite{Ahn2021} for a detailed discussion.}
	 	\label{fig:disorder}
	 \end{figure*}
 
	 The BdG Hamiltonians $\mathcal{H^\mathrm{eff}}$ and $\mathcal{H}_\mathrm{SE}^\mathrm{eff}$ will serve as the starting point for our numerical simulations presented in Sec.~\ref{sec:conductance}. (However, to demonstrate that it is reasonable to work with the effective low-energy model, we also present results obtained directly from the full multi-band Hamiltonian in the Appendix, showing that, in the regime considered in this work, the effective model captures all qualitative features of the full model.)  Throughout this work, we keep the parameters describing the Ge hole nanowire fixed since an extensive study of the pristine system has already been carried out in Ref.~\cite{Laubscher2023}. Explicitly, we fix the thickness of the quantum well to $L_z=22$~nm, which is a thickness that is routinely realized in current state-of-the-art experiments~\cite{Scappucci2021}. Other values of $L_z$ were investigated in Ref.~\cite{Laubscher2023} and lead only to quantitative, not qualitative changes to the topological phase diagrams. Furthermore, the width of the 1D channel to $L_y=15$~nm, the applied electric field to $\mathcal{E}=0.5$~V$/\mu$m, and the strain energy to $E_s=0.01$~eV~\footnote{The strain energy depends approximately linearly on the percentage of Si in the SiGe barrier. Reference~\cite{Sammak2019} reports $\varepsilon_{||}=-0.63\%$ in a Ge/SiGe quantum well with $20\%$ of Si in the barrier, from which we estimate $E_s=-|b|\varepsilon_0\approx 0.0237$~eV using $b\approx -2.16$~eV~\cite{Bir1974} and $\varepsilon_0\approx 1.74\varepsilon_{||}$~\cite{Terrazos2021}. Extrapolating the linear dependence, our choice of $E_s\approx0.01$~eV corresponds to $\approx$ 8.5\% of Si in the barrier, which is rather optimistic but not unrealistic compared to current state-of-the-art devices, where the Si content is typically 10\%-20\%.}.
	 
	 For a given choice of physical wire parameters, the corresponding effective parameters $\bar{m}$, $\bar{m}_s$, $g_\mathrm{eff}$, $\mu$, and $\alpha_{so}$ entering the low-energy model can be determined by fitting to the low-energy band structure of the full normal-state Hamiltonian $H_0$ [Eq.~(\ref{eq:normalH})], which we diagonalize taking the lowest 5 basis states of the square wells in the $y$ and $z$ direction into account (see also Appendix~\ref{app:parameters}). As an example, we show the (low-energy) spectrum of the full normal-state Hamiltonian $H_0$ and the spectrum of the effective Hamiltonian $H_0^\mathrm{eff}$, both at zero magnetic field, in Fig.~\ref{fig:setup}(b). At this point, it is worth noting that, since we are taking orbital effects into account [see Eq.~(\ref{eq:orb})], the effective parameters generally depend on the applied magnetic field. Since our main objective in Sec.~\ref{sec:conductance} will be to calculate tunneling conductance maps as functions of bias voltage and magnetic field, we will therefore work with $B$-dependent effective parameters, see the Appendix for a more detailed discussion.
	 
	 For the superconducting part of the Hamiltonian, we use parameters that roughly correspond to a Ge/Al hybrid nanowire unless specified otherwise. In particular, the critical field of the Al strip is taken to be $B_c=3$~T~\cite{Nichele2017,Suominen2017}. Furthermore, in the description without the self-energy term, we take the phenomenological proximity-induced gap at zero magnetic field to be $\Delta(0)=80~\mu$eV consistent with recent experiments~\cite{Vigneau2019,Valentini2024}. For these parameters, the bulk topological phase diagram of a Ge/Al hybrid nanowire is shown in Fig.~\ref{fig:setup}(c), and three examples of Majorana wave functions in the topological phase are shown in Fig.~\ref{fig:setup}(d). In the description with the self-energy term, we use $\Delta_0(0)=0.3$~meV for the superconducting gap of Al at zero magnetic field~\cite{Lutchyn2018} and $\gamma=0.1$~meV for the SC-SM coupling. At zero magnetic field, this again gives a proximity-induced gap of approximately $80~\mu$eV.

	\subsection{Disorder}
	\label{sec:model_disorder}
	We incorporate disorder into our effective low-energy model by replacing $\mu\rightarrow\mu-V_\mathrm{imp}(x)$, where $V_\mathrm{imp}(x)$ is a position-dependent effective disorder potential that phenomenologically describes the effect of randomly distributed onsite impurities in the proximitized Ge hole nanowire. Since the actual form of the effective disorder potential is not known in detail, we focus on two simple disorder models that allow us to study the main qualitative behavior of the system. First, we consider a random onsite disorder potential drawn from a Gaussian distribution with zero mean and standard deviation $\sigma$:
	\begin{equation}
	V_1(x)\sim \mathcal{N}(0,\sigma^2).\label{eq:disorder_uncorr}
	\end{equation}
	An example realization of $V_1(x)$ (one specific disorder configuration) is shown in Fig.~\ref{fig:disorder}(a). This form of disorder potential is frequently used throughout the literature on SC-SM hybrids, as it allows one to characterize the disorder by a single parameter $\sigma$ (the disorder strength). However, we note that $V_1(x)$ implicitly also depends on the lattice spacing $a$ of the discretized model (typically $a=5-10$~nm), which introduces an artificial length scale into the problem. Therefore, it is useful to compare the random Gaussian disorder model to a more physical effective disorder model suitable to describe disorder caused by random charged impurities in the limit of low to moderate impurity densities, which takes the form~\cite{Woods2021,Ahn2021} 
	\begin{equation}
	V_2(x)=V_0\sum_{j=1}^{N_\mathrm{imp}} (-1)^j\exp\left(-\frac{|x-x_j|}{\lambda}\right).\label{eq:disorder_corr}
	\end{equation}
	Here, $V_0$ is an average effective impurity amplitude, the $x_j$ for $j\in\{1,...,N_\mathrm{imp}\}$ denote the positions of the effective impurities, and $\lambda$ is the correlation length of the disorder. A given disorder configuration corresponds to $N_\mathrm{imp}$ impurity positions drawn from a uniform distribution that assigns an equal probability to all lattice sites in the discretized model. Note that, for simplicity, we choose a constant impurity amplitude $V_0$ for all impurities. More realistically, the impurity amplitudes could be drawn from a random distribution as well (e.g., from a Gaussian distribution centered around a mean impurity amplitude $\bar{V}_0$). An example realization of $V_2(x)$ (one specific disorder configuration) is shown in Fig.~\ref{fig:disorder}(b).
	
	To compare the two different disorder models on equal footing (and to assess to what extent the artificial random Gaussian potential can reasonably approximate certain aspects of an impurity-induced disorder potential), one can make use of the concept of `equivalent' disorder potentials introduced in Ref.~\cite{Ahn2021}. This equivalence relation is based on the observation that the high-frequency (short wavelength) components of the disorder potential do not significantly affect the Majorana physics in the hybrid SC-SM nanowire~\cite{Ahn2021}. To illustrate this concept, we define the Fourier transform $\tilde{V}$ of a given disorder potential $V$ as
	\begin{equation}
	\tilde{V}(k_n)=\frac{2a}{L_x}\sum_\ell V(a\ell)\sin(a\ell k_n)
	\end{equation}
	with $k_n=n\pi/L_x$ and where $\ell$ runs over all lattice sites. The disorder-averaged absolute value of the Fourier-transformed disorder potential, $\langle|\tilde{V}(k_n)|\rangle$, can then serve as a `spectral signature' characterizing a given type of disorder~\cite{Ahn2021}. As an example, in Fig.~\ref{fig:disorder}(c), we show the spectral signature of a random Gaussian disorder potential with strength $\sigma=50~\mu$eV on a lattice with lattice constant $a=5$~nm (red line), and compare it to the spectral signature of (1) a correlated disorder potential with impurity amplitude $V_0=80~\mu$eV, correlation length $\lambda=7$~nm, and impurity density $10/\mu$m (green line), and (2) a correlated disorder potential with impurity amplitude $V_0=50~\mu$eV, correlation length $\lambda=11$~nm, and impurity density $10/\mu$m (blue line). While the spectral signature of the Gaussian disorder potential is just a constant line, the spectral signatures of the correlated disorder potentials decay with increasing $k_n$ due to the presence of a finite correlation length. Nevertheless, all three disorder potentials have a comparable spectral signature at long wavelengths ($k_n< 0.05$~nm$^{-1}$). It was shown in Ref.~\cite{Ahn2021} that the long-wavelength components of the Fourier-transformed disorder potential determine how much the disorder affects the MZMs, which is why the three disorder models in our example can be considered approximately `equivalent' from the point of view of Majorana physics. However, for increasing correlation lengths, the approximation by a random Gaussian disorder potential becomes quantitatively less accurate (see for example Fig.~\ref{fig:disorder}(c), where the red line approximates the green line better than the blue line). In any case, in the parameter regime we consider in this work (relatively short correlation lengths $\lambda$), we find that the Gaussian disorder potential has a qualitatively similar effect on the tunneling conductance signatures of the Ge-based MZMs as an `equivalent' correlated disorder potential.
	
	Having discussed the disorder models that we will use in our simulations, we can further ask about the typical order of magnitude of the relevant parameters (in particular, of the disorder strength) in realistic Ge/SC hybrid devices. Since direct quantitative information on the disorder strength in hybrid Ge hole nanowires is not available, we base our estimate on the parent Ge quantum wells. Indeed, it is reasonable to assume that the typical disorder strength extracted from the Ge 2DHGs is a lower bound on the disorder strength in Ge-based SC-SM hybrid nanowires, because additional disorder will be introduced in the hybrid device by additional processing steps and the existence of interfaces. We can use the transport broadening $\Gamma=\hbar/2\tau$ as a simple measure of the disorder strength~\cite{Sau2012} in the best existing Ge 2DHGs, where $\tau=m^*\mu_{2D}/e$ is the scattering time, $m^*$ the effective in-plane mass, and $\mu_{2D}$ the 2D mobility. Assuming a hole mobility of $\mu_{2D}\sim 3.4\times 10^6$~cm$^{2}$/Vs~\cite{Stehouwer2023} and an in-plane hole mass of $m^*\sim 0.07m$, we find a very small value $\Gamma\sim 2.5~\mu$eV, indicating very weak disorder. Of course, the main caveat here is that, for MZM experiments, the Ge 2DHG has to be contacted with a superconductor, making it challenging to maintain the ultra-high mobilities of Ge holes in deeply buried wells. Encouragingly, however, the fabrication of shallow Ge quantum wells with high hole mobilities of $\sim 0.5\times 10^6$~cm$^{2}$/Vs has already been reported~\cite{Sammak2019}. Therefore, when we work with the random Gaussian disorder model, we will consider disorder strengths in the range $\sigma=5-500~\mu$eV, keeping in mind that a likely lower bound for the disorder strength in Ge-based hybrid nanowires fabricated from current state-of-the-art Ge 2DHGs is on the order of a couple of $\mu$eV.
	
	The situation is more complicated for the correlated disorder potential since the disorder now depends on three parameters (correlation length, impurity density, and impurity amplitude). At this point, we can only guess the order of magnitude of these parameters. (In principle, once the details of the experimental Ge-based hybrid device are known and experimental data becomes available, an analysis similar to Refs.~\cite{Woods2021,Ahn2021} could give a more accurate estimate.) In the following, we take the correlation length to be $\lambda=11$~nm, which is of the same order of magnitude as the correlation lengths estimated for InAs-based devices~\cite{Woods2021}. We consider different impurity amplitudes $V_0$ between $5~\mu$eV and $500~\mu$eV and different impurity densities $n_i$ in the range $2/\mu$m to $50/\mu$m.

	We note that while we only consider onsite potential disorder in this work, other parameters ($g$ factor, superconducting gap, strain, etc.) may exhibit random spatial fluctuations as well. The importance of these additional disorder mechanisms could potentially be assessed in a more detailed theoretical study once experimental transport data for Ge-based SC-SM hybrid nanowires becomes available. At the current stage, to keep the number of parameters reasonably low, we focus only on potential disorder, which is unavoidably present in any realistic device and has previously been shown to be the most significant type of disorder in the InAs-based Majorana platform. We also note that, since the disorder potential is added at the level of the effective low-energy two-band model, effects of disorder-induced inter-subband scattering are neglected in our description. However, this is not a severe limitation since the mobilities of the existing Ge 2DHGs are extremely high, such that we are working in the regime of weak disorder where the disorder strength is small compared to the typical inter-subband spacing.
	
	\begin{figure}[bt]
	\centering
	\includegraphics[width=0.8\columnwidth]{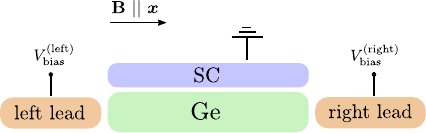}
	\caption{Schematic of the three-terminal device considered in this work. A Ge hole nanowire (green) is proximitized by a superconductor (blue) and a magnetic field is applied along the wire axis. Two semi-infinite normal leads (orange) are attached at the left and right end of the proximitized wire in order to measure the tunneling conductance. The superconductor is grounded.}
	\label{fig:threeterminal}
\end{figure}

	\subsection{Three-terminal device}
	\label{sec:model_three_terminal}
	
	To study the signatures of Ge-based MZMs in tunneling conductance measurements, we consider a three-terminal device as shown in Fig.~\ref{fig:threeterminal}, where two semi-infinite normal leads are attached at the right and left end of the proximitized Ge hole nanowire. The leads are described by the normal-state Hamiltonian $H_0^\mathrm{eff}$ [see Eq.~(\ref{eq:eff2band})] using the same parameters as in Sec.~\ref{sec:model_ge_eff} except for the chemical potential, which is shifted relative to the chemical potential $\mu$ of the Ge hole nanowire to $\mu-\mu_\mathrm{Lead}$ with $\mu_\mathrm{Lead}=-3$~meV. Furthermore, we also introduce tunnel barriers between the leads and the wire, which we model by a local chemical potential term at the first and last site of the discretized nanowire. The height of the barriers is fixed to $18$~meV throughout this work unless specified otherwise, since the qualitative effects of increasing/decreasing the barrier height are well known~\cite{Setiawan2017,Liu2017}. Finally, we also include a small dissipative term $i\Gamma$ with $\Gamma=10^{-4}$~meV in the Hamiltonian for the Ge hole nanowire~\cite{Liu2017,Liu2017b}.
	
	\subsection{Numerical method}
	\label{sec:numerical method}
	
The main objective of this work is to calculate the elements of the differential conductance matrix
	\begin{equation}
	G=\begin{pmatrix}G_{LL}&G_{LR}\\G_{RL}&G_{RR}\end{pmatrix}=\begin{pmatrix}\frac{\partial I_L}{\partial V_L}&-\frac{\partial I_L}{\partial V_R}\\-\frac{\partial I_R}{\partial V_L}&\frac{\partial I_R}{\partial V_R}\end{pmatrix},
	\end{equation}
	where $I_L$ ($I_R$) is the current entering the left (right) normal lead from the proximitized Ge hole nanowire, and $V_L$ ($V_R$) is the bias applied on the left (right) normal lead. Within the Blonder-Tinkham-Klapwijk (BTK) formalism~\cite{Blonder1982}, the elements of the conductance matrix are given by (here $i,j\in\{R,L\}$)
	\begin{align}
	G_{ii}&=\frac{e^2}{h}(N_i-T_{ii}^{N}+T_{ii}^{A}),\\
	G_{ij}&=\frac{e^2}{h}(T_{ij}^{N}-T_{ij}^{A}),\qquad i\neq j,
	\end{align}
	where $N_i$ is the number of channels in lead $i$ (here $N_i=2$) and $T_{ij}^{N}$ ($T_{ij}^{A}$) is the normal (anomalous) transmission amplitude from lead $j$ to lead $i$. To calculate these transmission coefficients numerically, we first discretize the effective one-dimensional Hamiltonian of the three-terminal device (see Sec.~\ref{sec:model_three_terminal}) on a finite lattice with lattice spacing $a=5$~nm. Then, we use the software package KWANT to calculate the transmission coefficients via the $S$ matrix formalism~\cite{Groth2014} for a discrete set of energies up to the proximity-induced superconducting gap, solving the scattering problem at each energy independently~\footnote{We note that the nonlinearity in $\omega$ appearing in the strong coupling limit does not further complicate the numerical procedure since we are evaluating the transmission coefficients for a discrete set of energies, such that, for each fixed energy, the Hamiltonian is a purely numerical matrix.}.

All our calculations are carried out at zero temperature $T=0$ since the typical temperatures at which MZM experiments are performed are very small ($T\sim 20$~mK), and finite temperature effects have already been studied in detail in the literature on InAs-based hybrid devices. Generally, as long as the topological gap is much larger than the typical temperature at which the experiment is performed (which is true for a wide range of parameters in our Ge system), a zero-temperature calculation is sufficient to obtain a qualitative understanding of the system.

	\section{Results}
	\label{sec:conductance}
	
	\begin{figure*}[bt]
		\centering
		\includegraphics[width=1\textwidth]{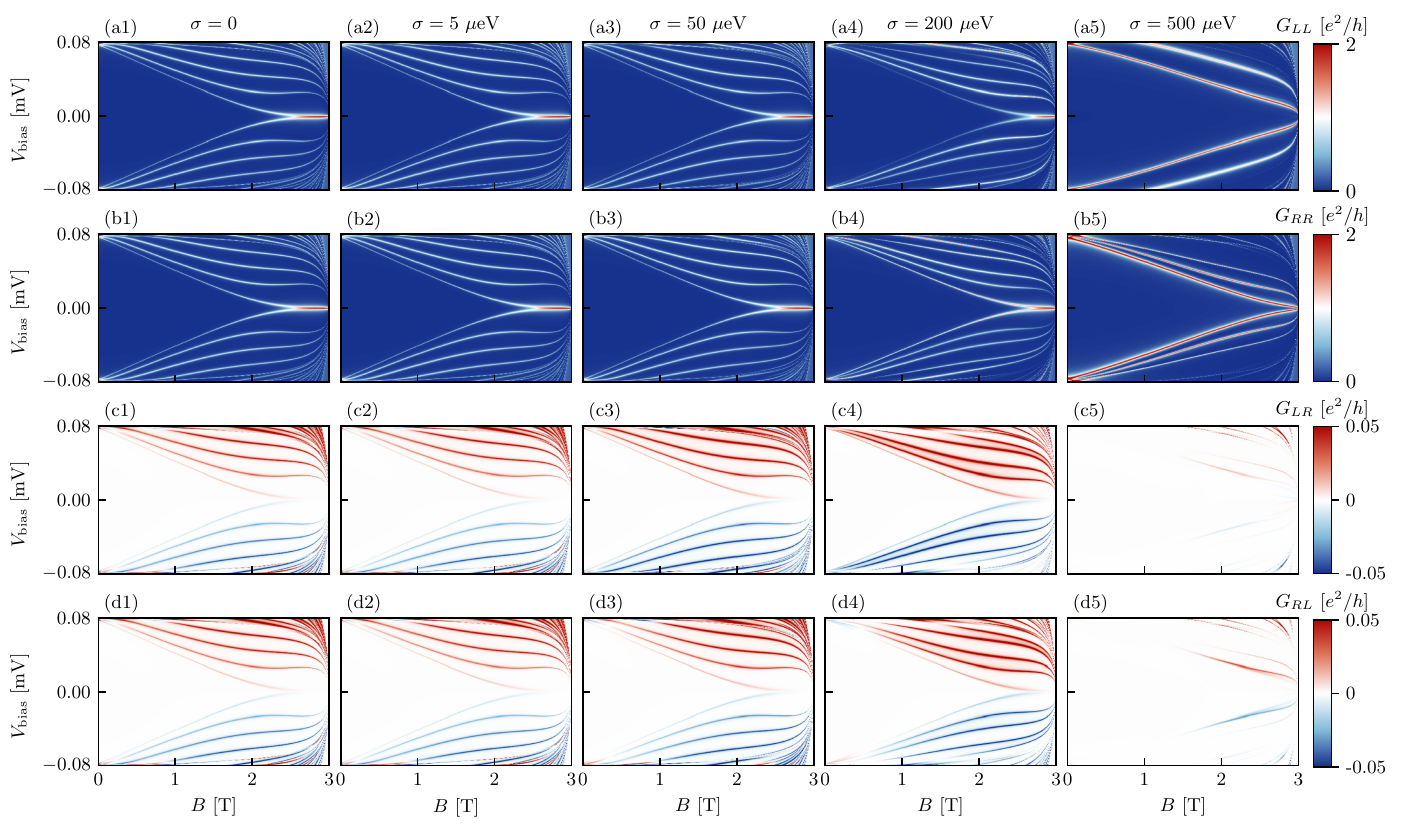}
		\caption{Local (first and second row) and nonlocal (third and fourth row) tunneling conductance spectra for a Ge/Al hybrid nanowire of length $L_x=1.5~\mu$m in the presence of random Gaussian disorder [see Eq.~(\ref{eq:disorder_uncorr})] for different disorder strengths $\sigma$. First column: $\sigma=0$, second column: $\sigma=5~\mu$eV, third column: $\sigma=50~\mu$eV, fourth column: $\sigma=200~\mu$eV, fifth column: $\sigma=500~\mu$eV.  For disorder strengths up to $\sim 200~\mu$eV, the local tunneling conductance spectra manifest MZM-induced ZBPs, and the nonlocal tunneling conductance spectra show a closing of the bulk gap at the topological phase transition. However, no clear bulk gap reopening can be observed since the size of the topological gap is comparable to the size of the finite-size induced level spacing. The Ge hole nanowire is modeled by Eq.~(\ref{eq:2band_self}) using effective parameters extracted from the full multiband Hamiltonian (see Sec.~\ref{sec:model_ge_full}) with  $L_y=15$~nm, $L_z=22$~nm, $\mathcal{E}=0.5$~V$/\mu$m, $E_s=0.01$~eV, and $\tilde\mu_0=0$. Furthermore, we use $\Delta_0(0)=0.3$~meV for the superconducting gap of Al at zero magnetic field, $B_c=3$~T for the critical field, and $\gamma=0.1$~meV for the SC-SM coupling.}
		\label{fig:conductance_onsite_L_1}
	\end{figure*}

\begin{figure*}[bt]
	\centering
	\includegraphics[width=1\textwidth]{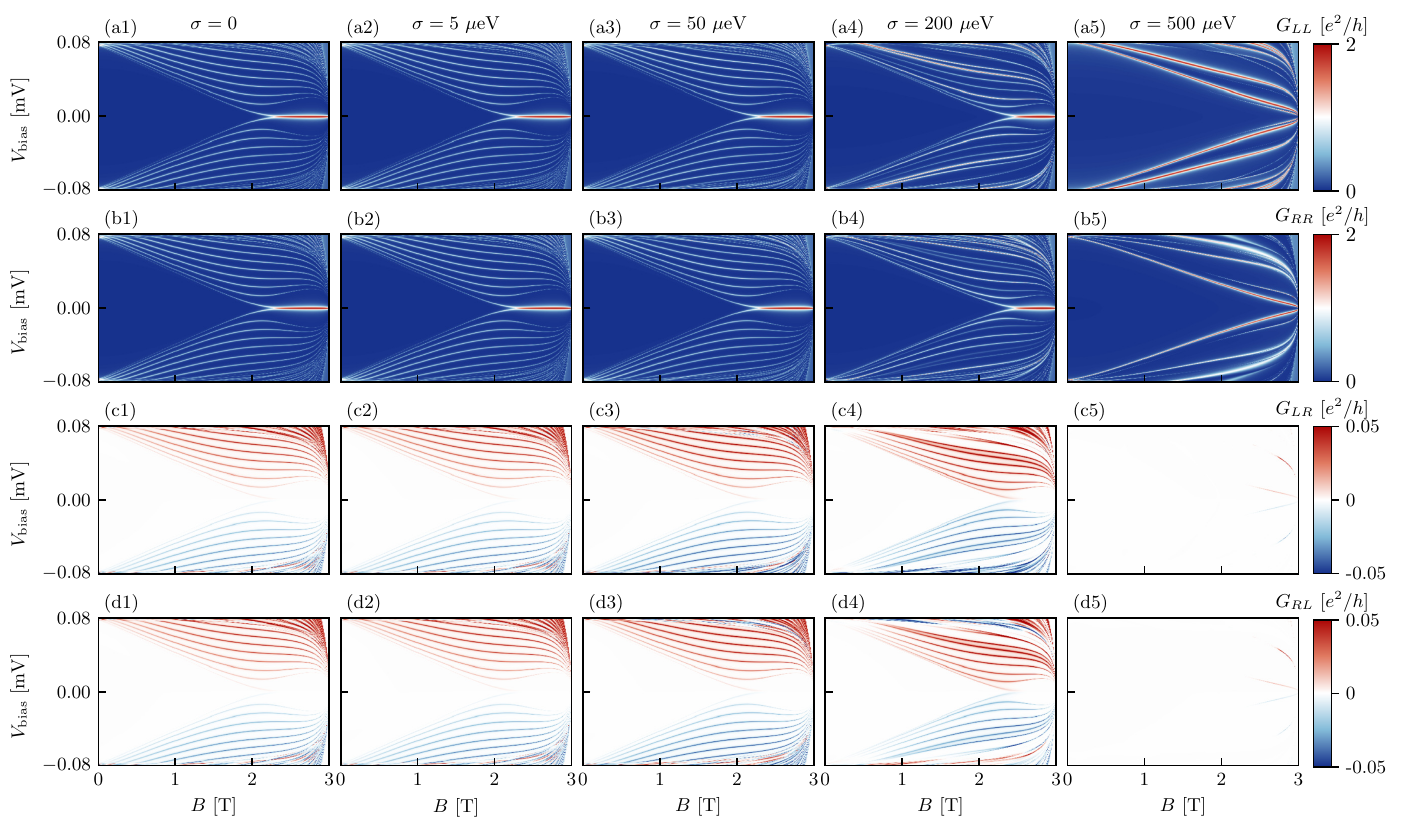}
	\caption{Local (first and second row) and nonlocal (third and fourth row) tunneling conductance spectra for a Ge/Al hybrid nanowire of length $L_x=3~\mu$m in the presence of random Gaussian disorder [see Eq.~(\ref{eq:disorder_uncorr})] for different disorder strengths $\sigma$. First column: $\sigma=0$, second column: $\sigma=5~\mu$eV, third column: $\sigma=50~\mu$eV, fourth column: $\sigma=200~\mu$eV, fifth column: $\sigma=500~\mu$eV. For disorder strengths up to $\sim 200~\mu$eV, the local tunneling conductance spectra manifest MZM-induced ZBPs, and the nonlocal tunneling conductance spectra show a closing and reopening of the bulk gap at the topological phase transition. The Ge hole nanowire is modeled by Eq.~(\ref{eq:2band_self}) using effective parameters extracted from the full multiband Hamiltonian (see Sec.~\ref{sec:model_ge_full}) with  $L_y=15$~nm, $L_z=22$~nm, $\mathcal{E}=0.5$~V$/\mu$m, $E_s=0.01$~eV, and $\tilde\mu_0=0$. Furthermore, we use $\Delta_0(0)=0.3$~meV for the superconducting gap of Al at zero magnetic field, $B_c=3$~T for the critical field, and $\gamma=0.1$~meV for the SC-SM coupling.}
	\label{fig:conductance_onsite_L_3}
\end{figure*}
	
	In this section, we present our numerical results for the local ($G_{LL}$ and $G_{RR}$)  and nonlocal ($G_{LR}$ and $G_{RL}$) tunneling conductance in Ge/SC hybrid nanowires as a function of bias voltage $V_\mathrm{bias}$ and magnetic field $B$ in the presence of disorder. Unless specified otherwise, all results are calculated for specific (randomly chosen) disorder configurations without any disorder averaging as appropriate for a specific experimental sample. In the main text, we focus on the case with the self-energy term [see Eq.~(\ref{eq:2band_self})] since the self-energy effect is non-negligible under real experimental conditions, while results without the self-energy term [see Eq.~(\ref{eq:2band_sc})] are presented in the Appendix for comparison. At this point, it is worth noting that the self-energy effect reduces the topological gap by a factor of $2-3$ compared to the case without the self-energy term, leading to a maximal topological gap of $E_\mathrm{gap}^\mathrm{max}\sim 20~\mu$eV ($\sim230$~mK) for our choice of wire parameters as compared to $E_\mathrm{gap}^\mathrm{max}\sim 50~\mu$eV if self-energy effects are neglected, see e.g. Fig.~\ref{fig:setup}(c). Nevertheless, since MZM experiments are typically carried out at very low temperatures $T\sim20$~mK, the topological gaps reported here should generally remain well-resolvable.
		
	We first present results for a random Gaussian disorder potential as introduced in Eq.~(\ref{eq:disorder_uncorr}). In Fig.~\ref{fig:conductance_onsite_L_1}, we show the differential tunneling conductance spectra for a Ge/Al hybrid nanowire of length $L_x=1.5~\mu$m for various different disorder strengths ranging from zero (pristine case) to $500~\mu$eV. Unless specified otherwise, the conductance spectra presented in this section are plotted using a color scale that ranges from $0$ to $2e^2/h$ for the local conductances and from $-0.05\,e^2/h$ to $+0.05\,e^2/h$ for the nonlocal conductances, with out-of-bound values being clipped.
	We start by discussing the pristine case shown in Figs.~\ref{fig:conductance_onsite_L_1}(a1-d1). Here, the local tunneling conductance spectra [see Figs.~\ref{fig:conductance_onsite_L_1}(a1,b1)] manifest end-to-end correlated MZM-induced ZBPs appearing at relatively large magnetic fields $B\gtrsim 2.5$~T. While the nominal topological phase transition occurs at a critical field of $B^*\sim 2$~T determined by the condition $V_{Z}^2(B^*)=\gamma^2+\mu^2$, the MZMs that emerge immediately after the transition have relatively long localization lengths compared to the wire length, such that the finite system considered here only manifests clear, quantized ZBPs at fields that significantly exceed $B^*$. Furthermore, the nonlocal conductance spectra [see Figs.~\ref{fig:conductance_onsite_L_1}(c1,d1)] show a closing of the bulk gap at the topological phase transition, with the conductance changing sign at zero bias as expected~\cite{Rosdahl2018}. However, the reopening of the bulk gap in the topological phase is not discernible here since the size of the topological gap is comparable to the size of the finite-size induced level spacing. (This will be different for longer nanowires and systems with larger topological gaps, see below.) Furthermore, we also mention that the nonlocal conductance is generally more than an order of magnitude smaller than the local conductance, meaning that measurements of relatively high precision ($\sim 10^{-2}\,e^2/h$) are necessary to resolve any of the nonlocal conductance signatures discussed in this work. Conductance measurements of precision up to $\sim 10^{-3}\,e^2/h$ have previously been reported in InAs-based devices~\cite{Menard2020,Puglia2021,Aghaee2022}.
	
	Next, in Figs.~\ref{fig:conductance_onsite_L_1}(a2-d2), we show the differential tunneling conductance spectra in the presence of weak disorder of strength $\sigma=5~\mu$eV. According to our estimates in Sec.~\ref{sec:model_disorder}, this approximately corresponds to the expected disorder strength in current state-of-the-art Ge 2DHGs. This value therefore represents a lower bound for the disorder strength in Ge-based SC-SM hybrid nanowires, where additional processing steps and the presence of interfaces introduce additional sources of disorder that are not present in deeply buried 2DHGs. We find that our simulated tunneling conductance maps look almost indistinguishable from the pristine case, reflecting the fact that MZMs are protected from weak disorder by virtue of their topological robustness. Increasing the disorder strength further, we show the differential tunneling conductance spectra for a disorder strength of $\sigma=50~\mu$eV in Figs.~\ref{fig:conductance_onsite_L_1}(a3-d3). Again, we find that the simulated conductance maps look almost indistinguishable from the pristine case even though the disorder strength already exceeds the topological gap. Deviations from the pristine case are eventually observed at a disorder strength of $\sigma=200~\mu$eV, see Figs.~\ref{fig:conductance_onsite_L_1}(a4-d4). The ZBPs in the local conductance [see Figs.~\ref{fig:conductance_onsite_L_1}(a4,b4)] remain clearly visible even in this case, but the topological phase transition seems to have moved to a slightly larger magnetic field as compared to the pristine case. Furthermore, the size of the maximal topological gap is reduced compared to the pristine case, showing that the stability of the MZMs decreases with increasing disorder strength. If the disorder strength is increased even further to $\sigma=500~\mu$eV, see Figs.~\ref{fig:conductance_onsite_L_1}(a5-d5), the topology of the system finally becomes suppressed due to disorder. The conductance spectra now generally depend strongly on the specific disorder configuration (see also below), with most samples neither exhibiting signatures of a topological phase transition in the nonlocal conductance [see Figs.~\ref{fig:conductance_onsite_L_1}(c5,d5)] nor MZM-induced ZBPs in the local conductance beyond that transition [see Figs.~\ref{fig:conductance_onsite_L_1}(a5,b5)]. Furthermore, depending on the specific disorder configuration, the left and right local conductance can now differ significantly and all end-to-end correlation is lost.

	Next, in Fig.~\ref{fig:conductance_onsite_L_3}, we present results for a longer nanowire of length $L_x=3~\mu$m, keeping all other parameters the same as before and again varying the disorder strength from zero (pristine case) to $500~\mu$eV. Due to the increased wire length, the topological transition  in the pristine case [see Figs.~\ref{fig:conductance_onsite_L_3}(a1-d1)] appears `sharper' compared to Figs.~\ref{fig:conductance_onsite_L_1}(a1-d1) in the sense that ZBPs appear almost immediately after the topological transition around $B^*\sim 2$~T. This is because the MZM localization length quickly becomes much shorter than the wire length, such that the MZMs at opposite ends of the wire do not significantly overlap even for magnetic fields that are only slightly larger than $B^*$. As such, the stability of the MZMs with respect to variations in $B$ is increased. Furthermore, due to the reduced finite-size level spacing, the bulk gap closing in the nonlocal conductance is now followed by a visible reopening (even though a high measurement precision $\sim 10^{-2}\,e^2/h$ is required for this closing and reopening transition to be resolvable in experiments). Apart from this, all of our qualitative findings regarding the effects of disorder remain the same as before, i.e., the MZM-induced ZBPs in the local conductance as well as the closing and reopening of the bulk gap in the nonlocal conductance remain intact up to a disorder strength of $\sim 200~\mu$eV. As the disorder strength is increased to $\sigma=500~\mu$eV, the topology of the system is destroyed and the local conductance spectra shown in Figs.~\ref{fig:conductance_onsite_L_3}(a5,b5) do no longer manifest ZBPs. Furthermore, the nonlocal conductance is now so strongly suppressed that Figs.~\ref{fig:conductance_onsite_L_3}(c5,d5) appear featureless for our choice of color scale.

We emphasize again that all results shown in this section are obtained for specific (randomly chosen) disorder configurations without any disorder averaging, as this is the relevant situation encountered in experiments. However, to ensure that the qualitative features discussed above are general rather than accidental, we have checked a large number of additional disorder configurations. In the Appendix, we show the corresponding disorder-averaged conductance maps for a hybrid Ge/Al nanowire of length $L_x=3~\mu$m for disorder strengths $\sigma=50$, $200$, $500~\mu$eV. For $\sigma=50~\mu$eV, we find that the disorder-averaged conductance spectra look almost indistinguishable from the pristine case. Furthermore, even for $\sigma=200~\mu$eV, the disorder-averaged conductance maps still exhibit all of the main qualitative features of the pristine case, with the local conductances manifesting well-discernible ZBPs in the topological phase, and the nonlocal conductances showing a closing and reopening of the bulk gap at a value close to the pristine critical field $B^*$. (We have also checked that, out of $150$ randomly chosen disorder configurations, none manifest a disorder-induced trivial ZBP in the trivial phase.) For $\sigma=500~\mu$eV, on the other hand, the conductance spectra depend strongly on the specific disorder configuration, with most samples neither exhibiting a topological phase transition nor ZBPs.

Our qualitative finding that the topological phase is stable up to disorder strengths that exceed the topological gap is consistent with previous studies of InAs-based hybrid devices~\cite{Pan2020,Pan2020b,Dassarma2021,Pan2021,Dassarma2023,Ahn2021,Dassarma2023b,Dassarma2023c}. However, the key difference between the InAs-based and the Ge-based platform is that the relevant energy scales (i.e., the size of the topological gap and the typical disorder strength) are likely very different. Indeed, a lower bound for the disorder strength in state-of-the-art InAs/Al hybrid devices can be estimated to be $\sim 0.6$~meV, leading to a maximal gap-to-disorder ratio of $\sim 1/3$. Consistent with this estimate, detailed theoretical studies~\cite{Dassarma2023b,Dassarma2023c} have shown that even the best InAs-based hybrid devices used in current MZM experiments~\cite{Aghaee2022} are still (at least) in the intermediate disorder regime. On the other hand, given our lower bound of only a couple of $\mu$eV for the disorder strength in the Ge 2DHG (see Sec.~\ref{sec:model_disorder}), a gap-to-disorder ratio of up to $\sim 5 $ could in principle be achieved in Ge-based devices. For such a low level of disorder, the local conductance maps look almost indistinguishable from the pristine case, see Figs.~\ref{fig:conductance_onsite_L_1}(a2-d2) and \ref{fig:conductance_onsite_L_3}(a2-d2). Even more importantly, while it is probably too optimistic to assume that the lower bound can actually be reached, Figs.~\ref{fig:conductance_onsite_L_1}(a3-d3) and \ref{fig:conductance_onsite_L_3}(a3-d3) show that even if the quality of the Ge-based hybrid device is an order of magnitude lower than the quality of the best Ge 2DHGs, the local conductance spectra still manifest clear ZBPs if and only if the system is in the topological phase, indicating that the system is in the weak disorder regime suitable for the realization and detection of topological MZMs.
	
		\begin{figure*}[bt]
		\centering
		\includegraphics[width=1\textwidth]{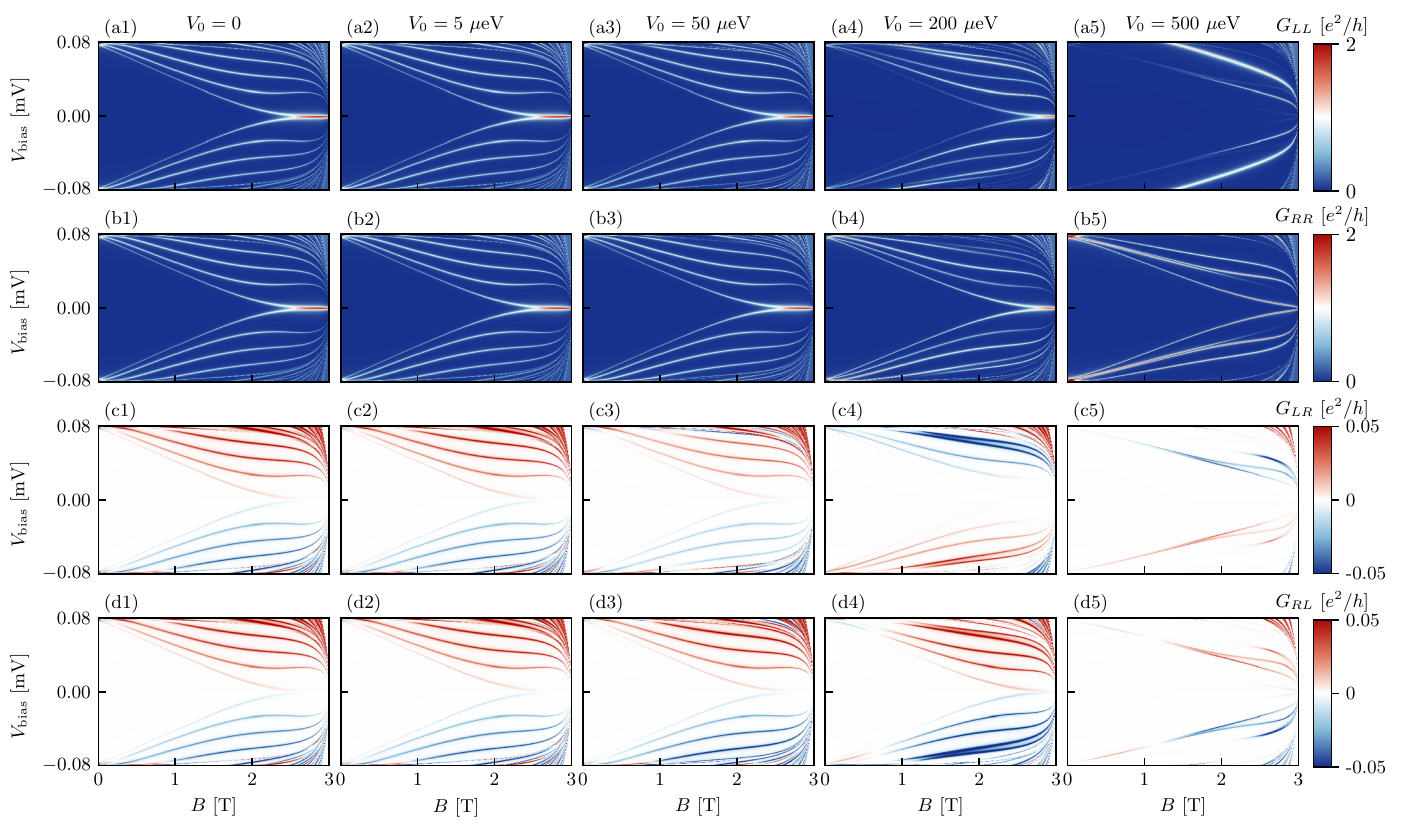}
		\caption{Local (first and second row) and nonlocal (third and fourth row) tunneling conductance spectra for a Ge/Al hybrid nanowire of length $L_x=1.5~\mu$m in the presence of correlated disorder [see Eq.~(\ref{eq:disorder_corr})] for different disorder strengths $V_0$. First column: $V_0=0$, second column: $V_0=5~\mu$eV, third column: $V_0=50~\mu$eV, fourth column: $V_0=200~\mu$eV, fifth column: $V_0=500~\mu$eV. For disorder strengths below $\sim 200~\mu$eV, the local tunneling conductance spectra manifest MZM-induced ZBPs, and the nonlocal tunneling conductance spectra show a closing of the bulk gap at the topological phase transition. However, no clear bulk gap reopening can be observed since the size of the topological gap is comparable to the size of the finite-size induced level spacing. The Ge hole nanowire is modeled by Eq.~(\ref{eq:2band_self}) using effective parameters extracted from the full multiband Hamiltonian (see Sec.~\ref{sec:model_ge_full}) with  $L_y=15$~nm, $L_z=22$~nm, $\mathcal{E}=0.5$~V$/\mu$m, $E_s=0.01$~eV, and $\tilde\mu_0=0$. Furthermore, we use $\Delta_0(0)=0.3$~meV for the superconducting gap of Al at zero magnetic field, $B_c=3$~T for the critical field, and $\gamma=0.1$~meV for the SC-SM coupling. For the disorder potential, we use the correlation length $\lambda=11$~nm and the impurity density $n_i=10/\mu$m.}
		\label{fig:conductance_correlated_L_1}
	\end{figure*}
	
	\begin{figure*}[bt]
		\centering
		\includegraphics[width=1\textwidth]{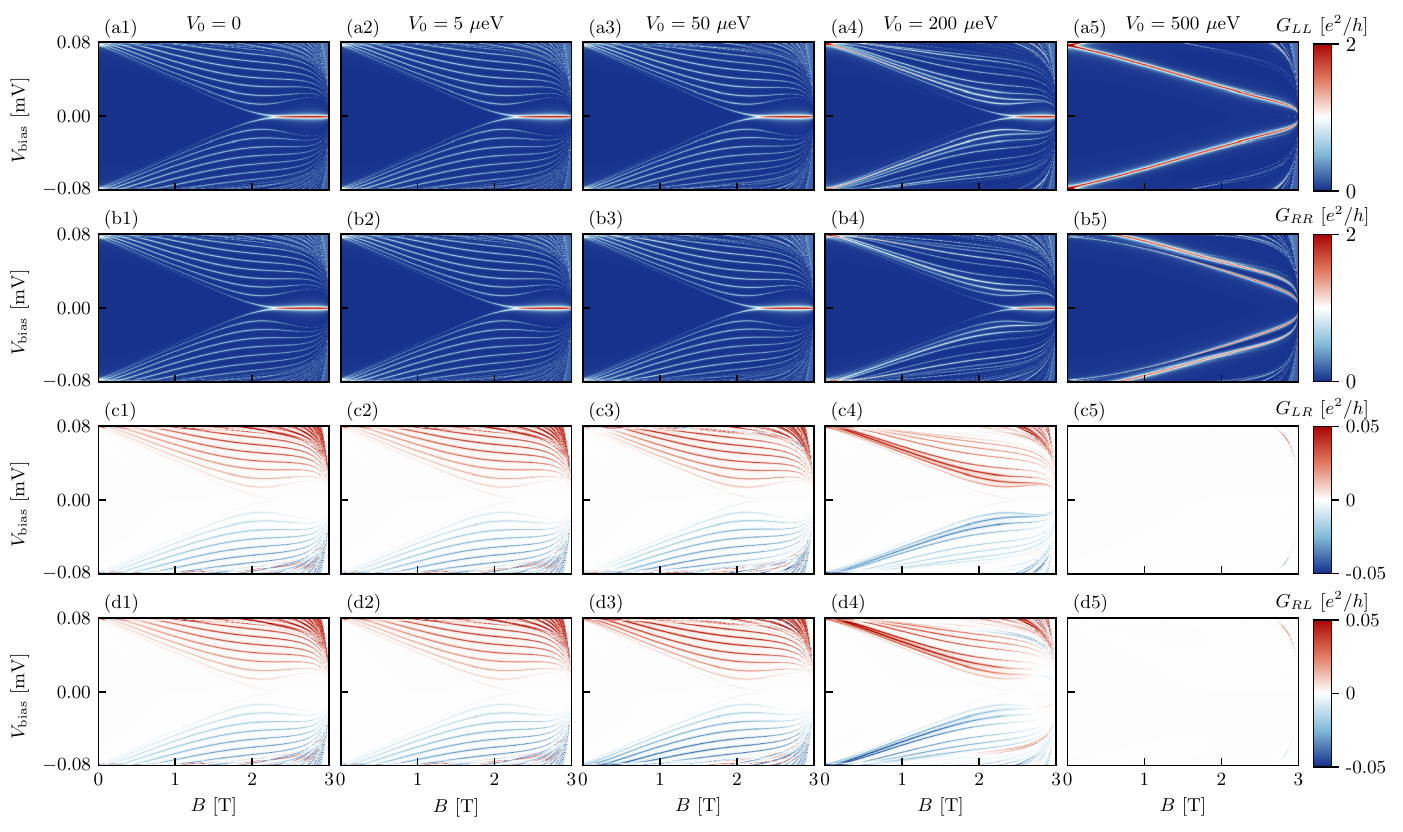}
		\caption{Local (first and second row) and nonlocal (third and fourth row) tunneling conductance spectra for a Ge/Al hybrid nanowire of length $L_x=3~\mu$m in the presence of correlated disorder [see Eq.~(\ref{eq:disorder_corr})] for different disorder strengths $V_0$. First column: $V_0=0$, second column: $V_0=5~\mu$eV, third column: $V_0=50~\mu$eV, fourth column: $V_0=200~\mu$eV, fifth column: $V_0=500~\mu$eV. For disorder strengths up to $\sim 200~\mu$eV, the local tunneling conductance spectra manifest MZM-induced ZBPs, and the nonlocal tunneling conductance spectra show a closing and reopening of the bulk gap at the topological phase transition. The Ge hole nanowire is modeled by Eq.~(\ref{eq:2band_self}) using effective parameters extracted from the full multiband Hamiltonian (see Sec.~\ref{sec:model_ge_full}) with  $L_y=15$~nm, $L_z=22$~nm, $\mathcal{E}=0.5$~V$/\mu$m, $E_s=0.01$~eV, and $\tilde\mu_0=0$. Furthermore, we use $\Delta_0(0)=0.3$~meV for the superconducting gap of Al at zero magnetic field, $B_c=3$~T for the critical field, and $\gamma=0.1$~meV for the SC-SM coupling. For the disorder potential, we use the correlation length $\lambda=11$~nm and the impurity density $n_i=10/\mu$m.}
		\label{fig:conductance_correlated_L_3}
	\end{figure*}
	
Next, in Fig.~\ref{fig:conductance_correlated_L_1}, we present local and nonlocal conductance maps for a Ge/Al hybrid nanowire of length $L_x=1.5~\mu$m in the presence of spatially correlated disorder as described by Eq.~(\ref{eq:disorder_corr}). Here, we fix the correlation length to $\lambda= 11$~nm and the impurity density to $10/\mu$m, while we vary the impurity amplitude $V_0$. (In the Appendix, we additionally show results for a fixed impurity amplitude but varying impurity densities). In Figs.~\ref{fig:conductance_correlated_L_1}(a1-d1), we show the tunneling conductance spectra in the pristine case. These panels are identical to the corresponding panels in Fig.~\ref{fig:conductance_onsite_L_1} and are just displayed here again for convenience. (In particular, we mention again that no bulk gap reopening is visible in the nonlocal conductance since the topological gap is comparable to the finite-size induced level spacing.) Furthermore, we find that the conductance maps for a disorder strength of $V_0=5~\mu$eV are indistinguishable from the pristine case, see Figs.~\ref{fig:conductance_correlated_L_1}(a2-d2). The same is still true when the disorder strength is increased by one order of magnitude to $V_0=50~\mu$eV, see Figs.~\ref{fig:conductance_correlated_L_1}(a3-d3). As the disorder strength is increased even further, deviations from the pristine case become visible, as can be seen from Figs.~\ref{fig:conductance_correlated_L_1}(a4-d4) for $V_0=200~\mu$eV. MZM-induced ZBPs can still be observed, but they only emerge at high magnetic fields close to the critical field of the superconductor, see Figs.~\ref{fig:conductance_correlated_L_1}(a4,b4). Furthermore, the two nonlocal conductances $G_{LR}$ and $G_{RL}$ can now significantly differ from each other (even in sign), see Figs.~\ref{fig:conductance_correlated_L_1}(c4,d4). Finally, Figs.~\ref{fig:conductance_correlated_L_1}(a5-d5) show that the topological superconductivity and the associated MZMs are clearly destroyed once the disorder strength reaches $V_0=500~\mu$eV.

In Fig.~\ref{fig:conductance_correlated_L_3}, we additionally present results for a longer nanowire of length $L_x=3~\mu$m in the presence of spatially correlated disorder, keeping all other parameters the same as in the previous paragraph. Comparing Figs.~\ref{fig:conductance_correlated_L_1} and \ref{fig:conductance_correlated_L_3} panel by panel, we find that the increased wire length leads to (i) an increased stability of the ZBPs with respect to variations of the magnetic field for all disorder strengths up to $V_0=200~\mu$eV, (ii) a visible bulk gap reopening after the topological transition for all disorder strengths up to $V_0=200~\mu$eV, and (iii) a stronger suppression of the nonlocal conductance for $V_0=500~\mu$eV. Other than that, all of our qualitative findings regarding the effects of disorder remain the same as before.

To compare Figs.~\ref{fig:conductance_onsite_L_1}--\ref{fig:conductance_onsite_L_3} (random Gaussian disorder) and Figs.~\ref{fig:conductance_correlated_L_1}--\ref{fig:conductance_correlated_L_3} (correlated disorder) on equal footing, we can make use of the concept of `equivalent' disorder potentials introduced in Ref.~\cite{Ahn2021} and briefly summarized in Sec.~\ref{sec:model_disorder}. Indeed, for our choice of disorder parameters (and our lattice constant $a=5$~nm), the disorder potential in the $n$th column of Fig.~\ref{fig:conductance_onsite_L_1} (Fig.~\ref{fig:conductance_onsite_L_3}) is approximately `equivalent' to the disorder potential in the $n$th column of Fig.~\ref{fig:conductance_correlated_L_1}  (Fig.~\ref{fig:conductance_correlated_L_3}) in the sense defined in Sec.~\ref{sec:model_disorder}, such that a column-by-column comparison of Figs.~\ref{fig:conductance_onsite_L_1}--\ref{fig:conductance_onsite_L_3} and \ref{fig:conductance_correlated_L_1}--\ref{fig:conductance_correlated_L_3} is meaningful. We observe the same qualitative behavior in Figs.~\ref{fig:conductance_onsite_L_1}--\ref{fig:conductance_onsite_L_3} and Figs.~\ref{fig:conductance_correlated_L_1}--\ref{fig:conductance_correlated_L_3}, namely, that the topological MZMs and the associated ZBPs remain intact up to the fourth column. This confirms that equivalent disorder potentials do indeed have a comparable effect on the low-energy Majorana physics of the Ge/SC hybrid nanowire, as was previously discussed in Ref.~\cite{Ahn2021} in the context of InAs-based devices. In particular, the comparison of Figs.~\ref{fig:conductance_onsite_L_1}--\ref{fig:conductance_onsite_L_3} and \ref{fig:conductance_correlated_L_1}--\ref{fig:conductance_correlated_L_3} shows that it is reasonable to work with a simple random Gaussian disorder potential in the parameter regime we consider. (Of course, since we are considering different disorder potentials and only display one specific disorder configuration, the actual conductance maps in the fourth and fifth column of Figs.~\ref{fig:conductance_onsite_L_1}--\ref{fig:conductance_onsite_L_3} look different from the corresponding conductance maps in the fourth and fifth column of Figs.~\ref{fig:conductance_correlated_L_1}--\ref{fig:conductance_correlated_L_3}, but in both cases the topology of the system remains intact up to disorder strengths $\sim 200~\mu$eV, which is the main feature we are interested in.)

	\begin{figure*}[bt]
	\centering
	\includegraphics[width=1\textwidth]{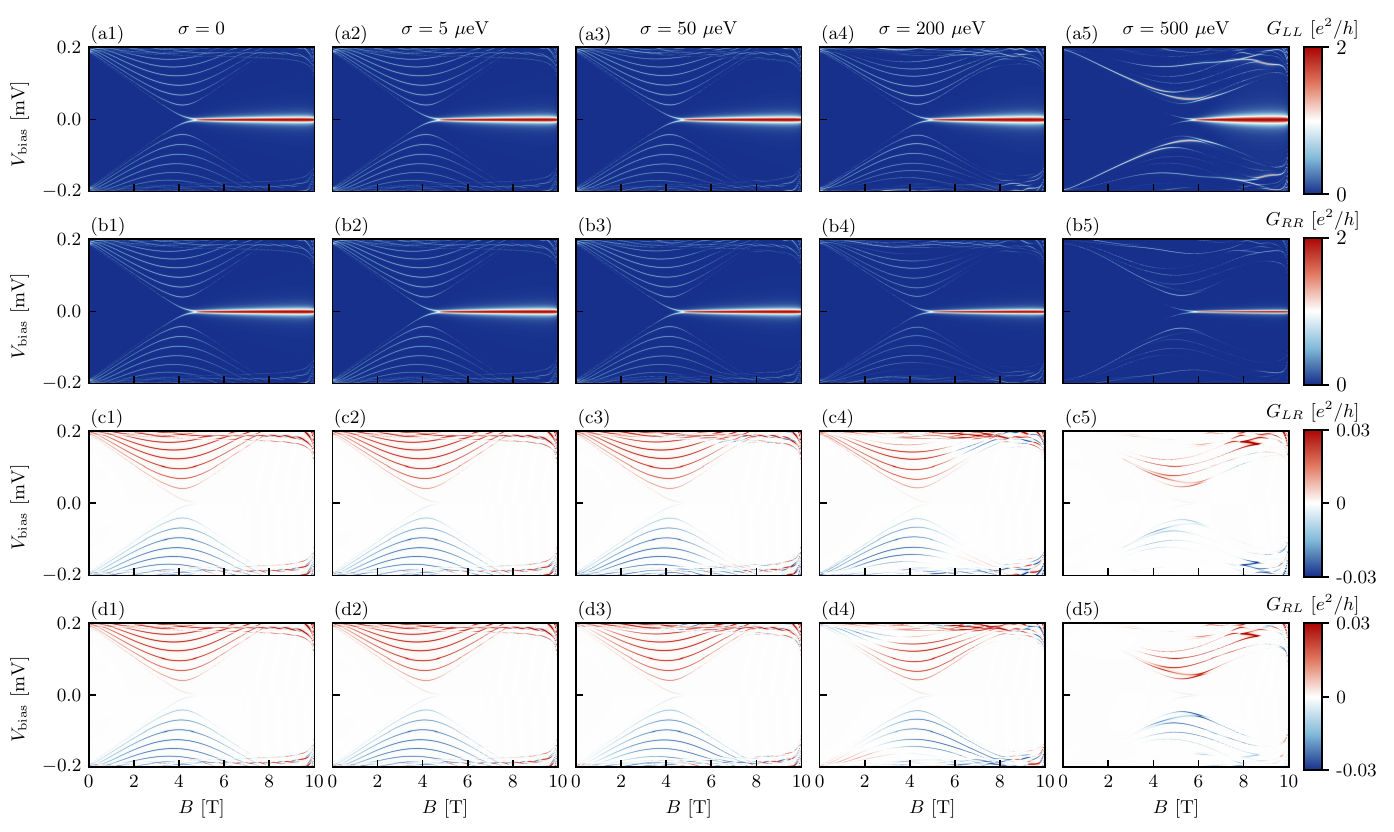}
	\caption{Local (first and second row) and nonlocal (third and fourth row) tunneling conductance spectra for a Ge/NbTiN hybrid nanowire of length $L_x=1.5~\mu$m in the presence of random Gaussian disorder [see Eq.~(\ref{eq:disorder_uncorr})] for different disorder strengths $\sigma$. First column: $\sigma=0$, second column: $\sigma=5~\mu$eV, third column: $\sigma=50~\mu$eV, fourth column: $\sigma=200~\mu$eV, fifth column: $\sigma=500~\mu$eV. For the entire range of disorder strengths considered here, the local tunneling conductance spectra manifest MZM-induced ZBPs, and the nonlocal tunneling conductance spectra show a clear closing and reopening of the bulk gap at the topological phase transition (however, for $\sigma=500~\mu$eV, the nonlocal conductance becomes very small, which is why Fig.~\ref{fig:conductance_NbTiN_zoom} shows the same plot again on a different color scale). The Ge hole nanowire is modeled by Eq.~(\ref{eq:2band_self}) using effective parameters extracted from the full multiband Hamiltonian (see Sec.~\ref{sec:model_ge_full}) with  $L_y=15$~nm, $L_z=22$~nm, $\mathcal{E}=0.5$~V$/\mu$m, $E_s=0.01$~eV, and $\tilde\mu_0=0$. Furthermore, we use $\Delta_0(0)=3$~meV for the superconducting gap of NbTiN at zero magnetic field, $B_c=10$~T for the critical field, and $\gamma=0.2$~meV for the SC-SM coupling. Due to the larger gap, we have also used a higher tunnel barrier of $30$~meV.}
	\label{fig:conductance_onsite_SE_NbTiN_L_1}
\end{figure*}

	\begin{figure*}[bt]
	\centering
	\includegraphics[width=1\textwidth]{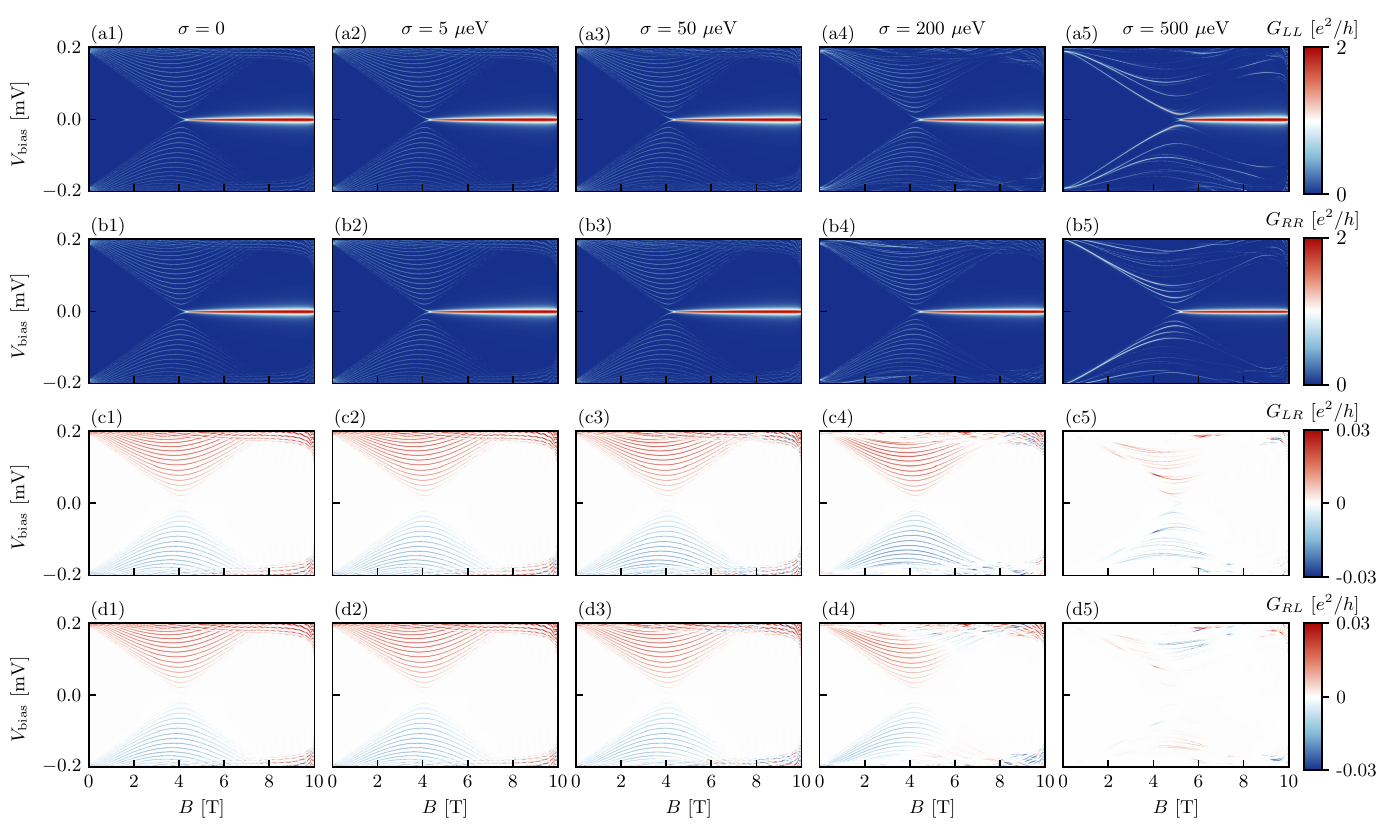}
	\caption{Local (first and second row) and nonlocal (third and fourth row) tunneling conductance spectra for a Ge/NbTiN hybrid nanowire of length $L_x=3~\mu$m in the presence of random Gaussian disorder [see Eq.~(\ref{eq:disorder_uncorr})] for different disorder strengths $\sigma$. First column: $\sigma=0$, second column: $\sigma=5~\mu$eV, third column: $\sigma=50~\mu$eV, fourth column: $\sigma=200~\mu$eV, fifth column: $\sigma=500~\mu$eV. For the entire range of disorder strengths considered here, the local tunneling conductance spectra manifest MZM-induced ZBPs, and the nonlocal tunneling conductance spectra show a clear closing and reopening of the bulk gap at the topological phase transition (however, for $\sigma=500~\mu$eV, the nonlocal conductance becomes very small, which is why Fig.~\ref{fig:conductance_NbTiN_zoom} shows the same plot again on a different color scale). The Ge hole nanowire is modeled by Eq.~(\ref{eq:2band_self}) using effective parameters extracted from the full multiband Hamiltonian (see Sec.~\ref{sec:model_ge_full}) with  $L_y=15$~nm, $L_z=22$~nm, $\mathcal{E}=0.5$~V$/\mu$m, $E_s=0.01$~eV, and $\tilde\mu_0=0$. Furthermore, we use $\Delta_0(0)=3$~meV for the superconducting gap of NbTiN at zero magnetic field, $B_c=10$~T for the critical field, and $\gamma=0.2$~meV for the SC-SM coupling. Due to the larger gap, we have also used a higher tunnel barrier of $30$~meV.}
	\label{fig:conductance_onsite_SE_NbTiN_L_3}
\end{figure*}
	
	In summary, Figs.~\ref{fig:conductance_onsite_L_1}--\ref{fig:conductance_correlated_L_3} show that, despite the relatively small pristine gaps, narrow Ge hole nanowires are an attractive Majorana platform  since the estimated disorder strength in current state-of-the-art Ge 2DHGs is only on the order of a couple of $\mu$eV. (We emphasize again that the important dimensionless quantity characterizing the quality of a Majorana platform is the gap-to-disorder ratio.) For a disorder strength of only a couple of $\mu$eV, the conductance maps are indistinguishable from the pristine case, see Figs.~\ref{fig:conductance_onsite_L_1}(a2-d2), \ref{fig:conductance_onsite_L_3}(a2-d2), \ref{fig:conductance_correlated_L_1}(a2-d2), and \ref{fig:conductance_correlated_L_3}(a2-d2). Of course, the actual disorder strength in a hybrid Ge/SC nanowire is certainly larger since additional disorder will be introduced into the system when the 1D channel is defined and when the system is proximitized by a superconductor. Nevertheless, Figs.~\ref{fig:conductance_onsite_L_1}(a3-d3), \ref{fig:conductance_onsite_L_3}(a3-d3), \ref{fig:conductance_correlated_L_1}(a3-d3), and \ref{fig:conductance_correlated_L_3}(a3-d3) show that even if the disorder strength exceeds the lower bound by an order of magnitude, the differential tunneling conductance maps remain almost indistinguishable from the pristine case. Therefore, one can reasonably hope that the weak disorder regime---where the topology of the system is intact, and the local conductance maps manifest well-discernible end-to-end correlated ZBPs as a signature of topological end MZMs---will be accessible in Ge-based hybrid devices.

	The main disadvantage of the Ge/Al hybrid devices studied in Figs.~\ref{fig:conductance_onsite_L_1}--\ref{fig:conductance_correlated_L_3} is the small effective $g$ factor of the Ge hole nanowire, which limits the pristine topological gaps to a maximum of $\sim 20~\mu$eV. This situation could be significantly improved if a superconductor with a larger bulk gap and/or a larger critical field than Al was used. A larger bulk gap would lead to a larger proximity-induced superconducting gap and, therefore, to a larger topological gap even in the regime of relatively weak SC-SM coupling (where self-energy effects are reduced). At the same time, a larger critical field would allow for larger magnetic fields to be applied without the superconducting gap collapsing, therefore leading to a larger topological phase space and increased topological gaps despite the smallness of the effective $g$ factor. To illustrate this, we now study the local and nonlocal tunneling conductance spectra in a Ge/NbTiN hybrid nanowire. A (soft) proximity-induced superconducting gap in a Ge/Si core/shell nanowire proximitized by NbTiN has been reported in Ref.~\cite{Su2016}. In our simulations, we set $\Delta_0(0)=3$~meV for the superconducting gap and $B_c=10$~T for the critical field of NbTiN~\cite{Lutchyn2018}. For the SC-SM coupling we choose $\gamma=0.2$~meV, which gives an induced gap of $\sim 0.2$~meV at zero field consistent with the value reported in Ref.~\cite{Su2016}. We note that this corresponds to the limit of very weak coupling to the superconductor $\gamma\ll\Delta_0(0)$, where the induced superconducting gap is approximately equal to $\gamma$.
	
	In Figs.~\ref{fig:conductance_onsite_SE_NbTiN_L_1} and \ref{fig:conductance_onsite_SE_NbTiN_L_3}, we present results for Ge/NbTiN hybrid nanowires of length $L_x=1.5~\mu$m and $L_x=3~\mu$m, respectively. We note that we now use a slightly different color scale for the nonlocal conductance, with the color bar ranging from $-0.03\,e^2/h$ to $+0.03\,e^2/h$. In Figs.~\ref{fig:conductance_onsite_SE_NbTiN_L_1}(a1-d1) and \ref{fig:conductance_onsite_SE_NbTiN_L_3}(a1-d1), we show the differential tunneling conductance spectra in the pristine case. Here, the topological phase transition happens at a larger magnetic field $B^*\sim 4$~T than in the Ge/Al device (see Figs.~\ref{fig:conductance_onsite_L_1}-\ref{fig:conductance_correlated_L_3}) because we are working with a larger SC-SM coupling $\gamma$. At the same time, due to the larger superconducting gap and critical field of NbTiN (and the larger SC-SM coupling), the topological gap is enhanced significantly compared to Figs.~\ref{fig:conductance_onsite_L_1}-\ref{fig:conductance_correlated_L_3}, with the maximal topological gap almost reaching $200~\mu$eV. In Figs.~\ref{fig:conductance_onsite_SE_NbTiN_L_1}(a2-d2) and \ref{fig:conductance_onsite_SE_NbTiN_L_3}(a2-d2), we show the tunneling conductance spectra at a disorder strength of $\sigma=5~\mu$eV, and in Figs.~\ref{fig:conductance_onsite_SE_NbTiN_L_1}(a3-d3) and \ref{fig:conductance_onsite_SE_NbTiN_L_3}(a3-d3) the same quantities are shown for a disorder strength of $\sigma=50~\mu$eV. In both cases, no deviations from the pristine case are visible by eye. In particular, for both wire lengths, the local conductance spectra manifest clear end-to-end correlated ZBPs whenever the system is in the topological phase. Furthermore, the nonlocal conductances show a clear gap closing and reopening transition as the system enters the topological phase.
	The same is still true in Figs.~\ref{fig:conductance_onsite_SE_NbTiN_L_1}(a4-d4) and \ref{fig:conductance_onsite_SE_NbTiN_L_3}(a4-d4) for a disorder strength of $\sigma=200~\mu$eV, even though the nonlocal conductance signatures are now slightly weaker than in the pristine case. Finally, Figs.~\ref{fig:conductance_onsite_SE_NbTiN_L_1}(a5-d5) and \ref{fig:conductance_onsite_SE_NbTiN_L_3}(a5-d5) show the tunneling conductance spectra for a disorder strength of $\sigma=500~\mu$eV, where deviations from the pristine case become visible. Nevertheless, one can still observe clear ZBPs in the local conductance, see Figs.~\ref{fig:conductance_onsite_SE_NbTiN_L_1}(a5,b5) and \ref{fig:conductance_onsite_SE_NbTiN_L_3}(a5,b5). The nonlocal conductance is reduced compared to the pristine case (especially in the longer nanowire with $L_x=3~\mu$m), but a gap closing and reopening can still be observed if the precision of the nonlocal conductance measurement is high enough ($\sim 10^{-3}e^2/h$). To demonstrate this, we show an alternative version of the plots corresponding to Figs.~\ref{fig:conductance_onsite_SE_NbTiN_L_1}(c5,d5) and \ref{fig:conductance_onsite_SE_NbTiN_L_3}(c5,d5) in the Appendix, where we have restricted the color bar to smaller values in the range $-0.005\, e^2/h$ to $0.005\, e^2/h$.
	
	Of course, a substantial amount of additional experimental work would be required to confirm whether the realization of MZMs in Ge/NbTiN hybrid devices is a feasible idea. (One of the problems with NbTiN is that so far only relatively soft proximity-induced gaps with significant residual subgap conductance have been observed.) As such, Figs.~\ref{fig:conductance_onsite_SE_NbTiN_L_1} and \ref{fig:conductance_onsite_SE_NbTiN_L_3} should only be seen as a simple example that allows us to illustrate the significant advantages of a superconductor with a larger critical field and/or a larger gap. Finding suitable superconductors for use in Ge/SC hybrid devices is a topic of significant current interest in experimental physics and materials sciences, and promising first results have already been obtained. For example, a hard proximity-induced gap has recently been observed in Ge/PtSiGe hybrid devices~\cite{Tosato2022}. Furthermore, recent works have reported the successful growth of tantalum germanide~\cite{Strohbeen2023b} and hyperdoped Ge~\cite{Strohbeen2023} on Ge substrates, with the former exhibiting a large in-plane critical field $B_c\sim 5$~T. Our theoretical conductance maps in Figs.~\ref{fig:conductance_onsite_SE_NbTiN_L_1} and \ref{fig:conductance_onsite_SE_NbTiN_L_3} show that, if experimentally feasible, hybrid Ge/SC devices with large critical fields constitute an ideal MZM platform with large pristine topological gaps and low intrinsic levels of disorder.

\section{Conclusions}
\label{sec:conclusions}

We have theoretically studied the local and nonlocal tunneling conductance spectra for Ge-based Majorana nanowires taking into account the effects of unintentional random disorder in the chemical potential. We have presented results for two different wire lengths, two different disorder models (random Gaussian disorder and spatially correlated disorder), two different parent superconductors (Al and NbTiN), and various disorder strengths, allowing us to critically discuss the stability of the experimental Majorana signatures in the local and nonlocal conductance in the presence of disorder. Based on experimentally reported peak mobilities, we have estimated the disorder strength in current state-of-the-art Ge 2DHGs to be on the order of a couple of $\mu$eV, which provides a natural lower bound for the disorder strength in Ge-based hybrid devices. Our numerical simulations for Ge/Al hybrid devices show that even if this lower bound is exceeded by an order of magnitude (due to additional disorder introduced by additional processing steps and/or interfaces in the hybrid device), the local conductance spectra remain almost indistinguishable from the pristine case. Our work suggests that the weak disorder regime---where the topology of the system is intact, and the local conductance maps manifest well-discernible end-to-end correlated ZBPs if and only if the system hosts topological end MZMs---could be accessible in realistic Ge/SC hybrid nanowires fabricated from current state-of-the-art Ge 2DHGs. This should be compared to the current situation in InAs/Al hybrid devices, where, after more than 10 years of intense experimental efforts, the disorder strength in the parent 2DEG is still at least $\sim 3$ times larger than the pristine topological gap, meaning that even the best InAs-based hybrid devices that are currently available are still in the intermediate disorder regime~\cite{Dassarma2023b,Dassarma2023c}.

While the small in-plane $g$ factor of the Ge hole nanowire limits the topological gaps in Ge/Al devices to a few tens of $\mu$eV, our simulations demonstrate that the situation can be improved if a superconductor with a larger critical field and/or larger gap than Al is used. Using parameters that roughly correspond to a Ge/NbTiN hybrid device, we show that, in this case, the experimental Majorana signatures remain stable even if the quality of the hybrid system is reduced by two orders of magnitude compared to the quality of the best currently available Ge 2DHGs. These results can be seen as a motivation for future experimental work toward the fabrication of Ge/SC hybrids using superconductors other than Al.

\begin{figure*}[bt]
	\centering
	\includegraphics[width=1\textwidth]{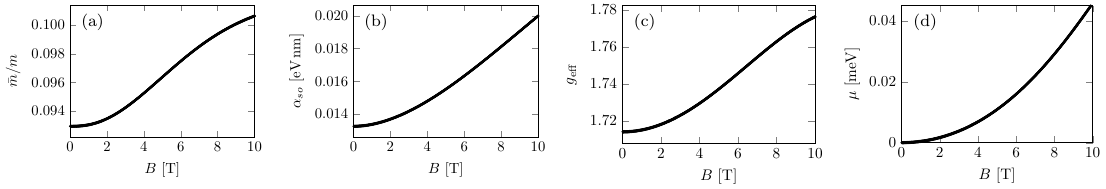}
	\caption{Effective parameters entering Eqs.~(\ref{eq:eff2band}), (\ref{eq:2band_sc}), and (\ref{eq:2band_self}) as a function of magnetic field $B$. (a) Effective mass $\bar{m}$. (b) Effective spin-orbit coupling $\alpha_{so}$. (c) Effective $g$ factor $g_{\mathrm{eff}}$. (d) Effective chemical potential $\mu$ (measured from the spin-orbit point).}
	\label{fig:eff_params}
\end{figure*}

Our work establishes that the advantages of the Ge hole-based hybrid Majorana platforms, namely, their large spin-orbit coupling strength and low disorder (i.e. high mobility), makes them attractive as topological quantum computing platforms. We also mention another advantage, which is that one can use 2D Ge hole systems for scaling up from a single nanowire to a large number of nanowires on a single 2D device, creating a gate-tunable multiqubit system similar to that proposed in the InAs/Al Majorana platform. On the other hand, the main disadvantage of the Ge platform is the small $g$ factor, which limits both the topological phase space and the maximal topological gaps that can be achieved. This situation could be improved by careful band structure engineering and optimization of the device geometry, which might make it possible to increase the effective $g$ factors of Ge hole nanowires beyond the values that we have considered here. Additionally, the geometry of the superconducting thin film could be optimized to increase the critical field for a given superconductor. Apart from the $g$ factor, the topological gap depends on parameters such as $\Delta_{3/2}/\Delta_{1/2}$ and the interplay of strain and disorder in a way that would be interesting to study in future theoretical work. In the end, the feasibility of Ge holes as a Majorana platform can only be decisively established through experimental work; all our theory shows is that such work may be warranted given the slow progress and the disorder problems plaguing the InAs/Al Majorana platform~\cite{Aghaee2022,Dassarma2023b,Dassarma2023c}.

Finally, we note that, while we have focused on proximitized gate-defined Ge hole nanowires in this work, our message regarding the exceptional materials quality of the Ge platform is also relevant in the context of other Ge-based hybrid devices that could potentially host MZMs, such as, e.g., Ge-based planar Josephson junctions~\cite{Luethi2022,Luethi2023}.

\section*{Acknowledgments}
This work is supported by the Laboratory for Physical Sciences through the Condensed Matter Theory Center.

\appendix

\section{Effective parameters}
\label{app:parameters}

The low-energy spectrum of the gate-defined Ge hole nanowire is obtained from the full multiband Hamiltonian $H=H_0+H_{sc}$ in BdG form by expanding its eigenstates in terms of suitable basis functions that solve the confinement problem~\cite{Csontos2009,Kloeffel2011,Kloeffel2018,Adelsberger2021,Adelsberger2022,Milivojevic2021}. Assuming translational invariance along the channel for now, we can write the spatially varying part of these basis functions as $\varphi_{k_x,p,q}(x,y,z)=e^{ik_xx}\varphi_{p}(y)\varphi_{q}(z)$, where
\begin{align}
\varphi_{p}(y)&=\begin{cases} \sqrt{2/L_y}\sin\left(p\pi (\frac{y}{L_y}+\frac{1}{2})\right)&|y|<L_y/2,\\0&\mathrm{ otherwise, }\end{cases}\\
\varphi_{q}(z)&=\begin{cases} \sqrt{2/L_z}\sin\left(q\pi (\frac{z}{L_z}+\frac{1}{2})\right)&|z|<L_z/2,\\0&\mathrm{ otherwise, }\end{cases}
\end{align}
are the eigenfunctions of the infinite square well along the $y$ and $z$ direction, respectively (here $p,q\in\{1,2,...\}$). For our numerical simulations, we project the full BdG Hamiltonian into the subspace spanned by the first 5 basis functions $p,q\in\{1,2,...,5\}$ for each spatial direction, which results in a $200\times200$ effective Hamiltonian that can be diagonalized numerically.

By fitting to the bands of the full multiband Hamiltonian as a function of $k_x$ for small $k_x$, we then determine the parameters entering the effective two-band Hamiltonian given in Eq.~(\ref{eq:eff2band}). In Fig.~\ref{fig:eff_params}, we show the effective mass $\bar{m}$, the effective spin-orbit coupling $\alpha_{so}$, the effective chemical potential $\mu$ measured from the spin-orbit point of the lowest confinement-induced subband, and the effective $g$ factor $g_\mathrm{eff}$ extracted from the full multiband Hamiltonian as functions of $B$. For our choice of wire parameters and the range of energies we are interested in, we find that the effective spin-dependent mass $\bar{m}_s$ is very small, which is why we will neglect it here. Since orbital effects are taken into account in our description of the Ge hole nanowire, all effective parameters generally depend on $B$, even though this dependence is relatively weak for our choice of wire parameters. However, note the change in effective chemical potential $\mu$ with increasing $B$, which is an effect that is well-known also for electron nanowires~\cite{Dmytruk2018}.

\section{Comparison between the full model and the effective model}
\label{app:comparison}

\begin{figure*}[bt]
	\centering
	\includegraphics[width=1\textwidth]{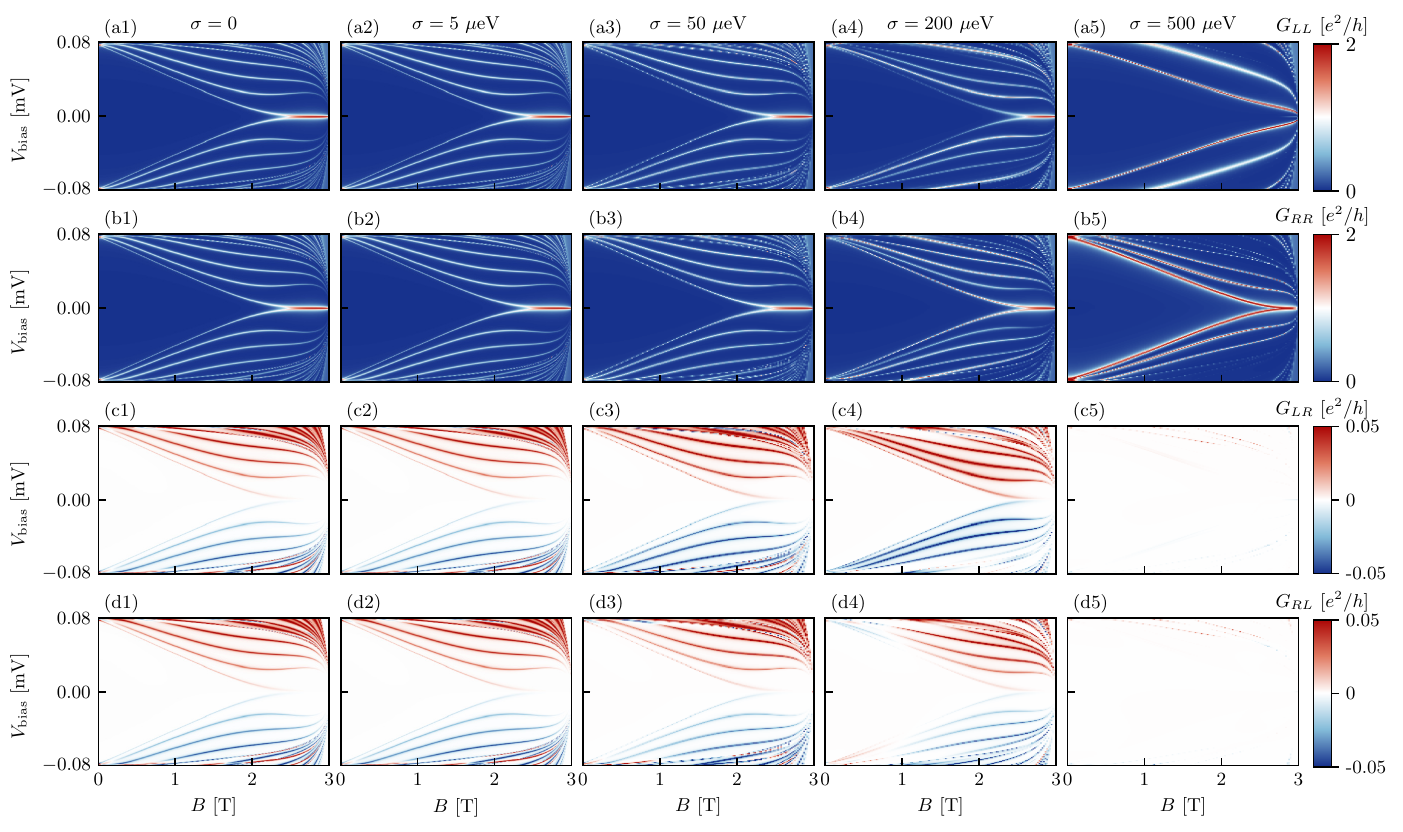}
	\caption{Local (first and second row) and nonlocal (third and fourth row) tunneling conductance spectra for a Ge/Al hybrid nanowire of length $L_x=1.5~\mu$m obtained from the full multiband Hamiltonian in the presence of random Gaussian disorder [see Eq.~(\ref{eq:disorder_uncorr})] for different disorder strengths $\sigma$. First column: $\sigma=0$, second column: $\sigma=5~\mu$eV, third column: $\sigma=50~\mu$eV, fourth column: $\sigma=200~\mu$eV, fifth column: $\sigma=500~\mu$eV. The parameters for the Ge hole nanowire are $L_y=15$~nm, $L_z=22$~nm, $\mathcal{E}=0.5$~V$/\mu$m, $E_s=0.01$~eV, and $\tilde\mu_0=0$. Furthermore, we use $\Delta_0(0)=0.3$~meV for the superconducting gap of Al at zero magnetic field, $B_c=3$~T for the critical field, and $\gamma=0.1$~meV for the SC-SM coupling.}
	\label{fig:conductance_comparison}
\end{figure*}

In Fig.~\ref{fig:conductance_comparison}, we show the conductance spectra obtained from the full multiband model for a Ge/Al hybrid nanowire of length $L_x=1.5~\mu$m in the presence of random Gaussian onsite disorder using the same specific disorder configuration as in Fig.~\ref{fig:conductance_onsite_L_1}. Comparing Figs.~\ref{fig:conductance_comparison} and \ref{fig:conductance_onsite_L_1} panel by panel, we see that the conductance maps calculated from the effective model look qualitatively very similar to the conductance maps calculated from the full model, justifying the use of the effective model in the main text. In the full model, we have worked with an effective disorder potential that is constant throughout the cross section of the wire and varies as a function of $x$ according to  Eq.~(\ref{eq:disorder_uncorr}), such that the disorder again effectively corresponds to random variations in the chemical potential. Of course, a more realistic disorder potential should depend on all three spatial coordinates, but, since no information about the actual disorder potential in experimental Ge/Al hybrid nanowire samples is currently available, not much would be gained by using a more complicated disorder model at this point.

\section{Results without the self-energy term}
\label{app:no_SE}

\begin{figure*}[bt]
	\centering
	\includegraphics[width=1\textwidth]{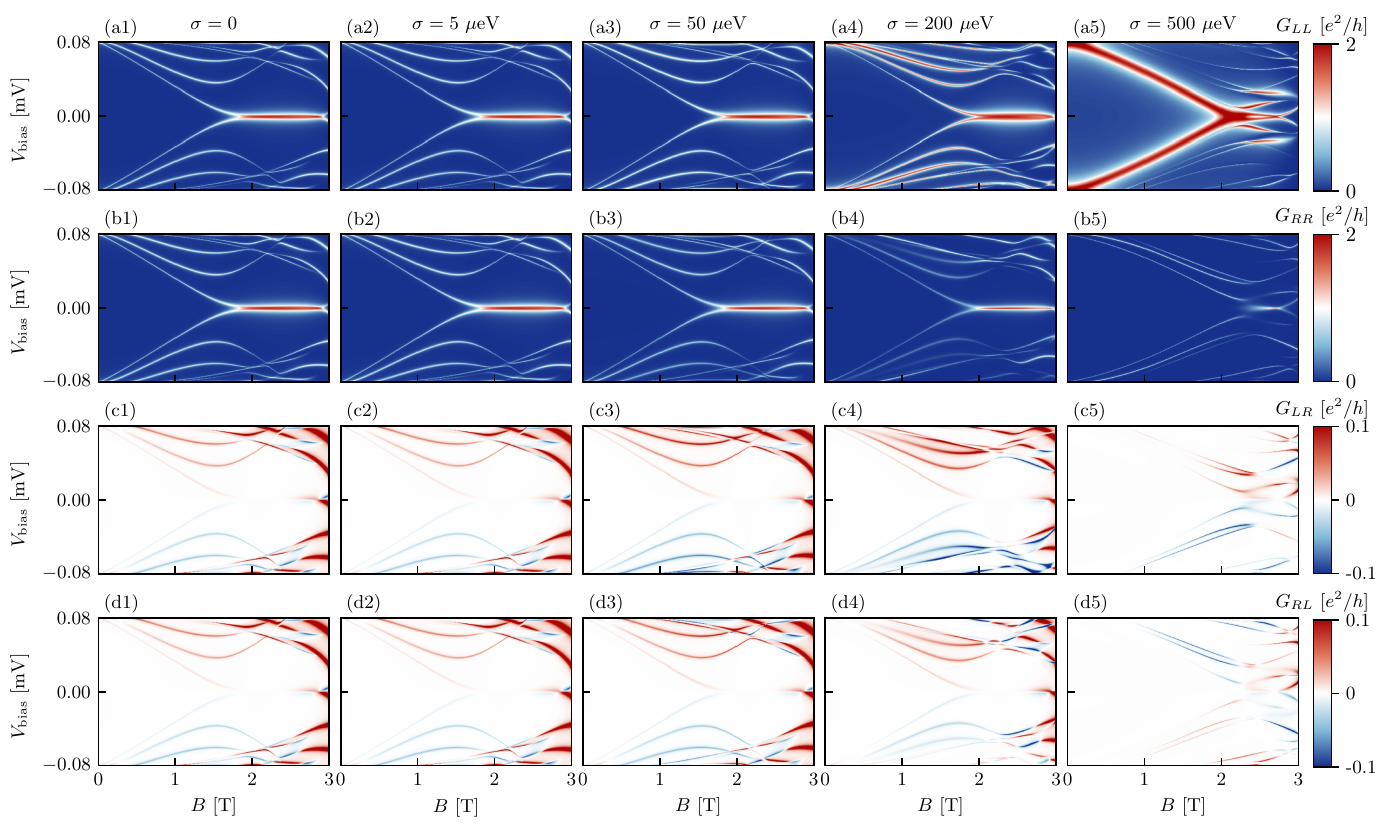}
	\caption{Local (first and second row) and nonlocal (third and fourth row) tunneling conductance spectra for a Ge/Al hybrid nanowire of length $L_x=1.5~\mu$m in the presence of random Gaussian disorder [see Eq.~(\ref{eq:disorder_uncorr})] for different disorder strengths $\sigma$. First column: $\sigma=0$, second column: $\sigma=5~\mu$eV, third column: $\sigma=50~\mu$eV, fourth column: $\sigma=200~\mu$eV, fifth column: $\sigma=500~\mu$eV. The Ge hole nanowire is modeled by Eq.~(\ref{eq:2band_sc}) using effective parameters extracted from the full multiband Hamiltonian (see Sec.~\ref{sec:model_ge_full}) with  $L_y=15$~nm, $L_z=22$~nm, $\mathcal{E}=0.5$~V$/\mu$m, $E_s=0.01$~eV, and $\tilde\mu_0=0$. The proximity-induced superconducting gap at zero magnetic field is set to $\Delta(0)=80~\mu$eV, and the critical field is taken to be $B_c=3$~T.}
	\label{fig:conductance_without_SE_L_1}
\end{figure*}

\begin{figure*}[bt]
	\centering
	\includegraphics[width=1\textwidth]{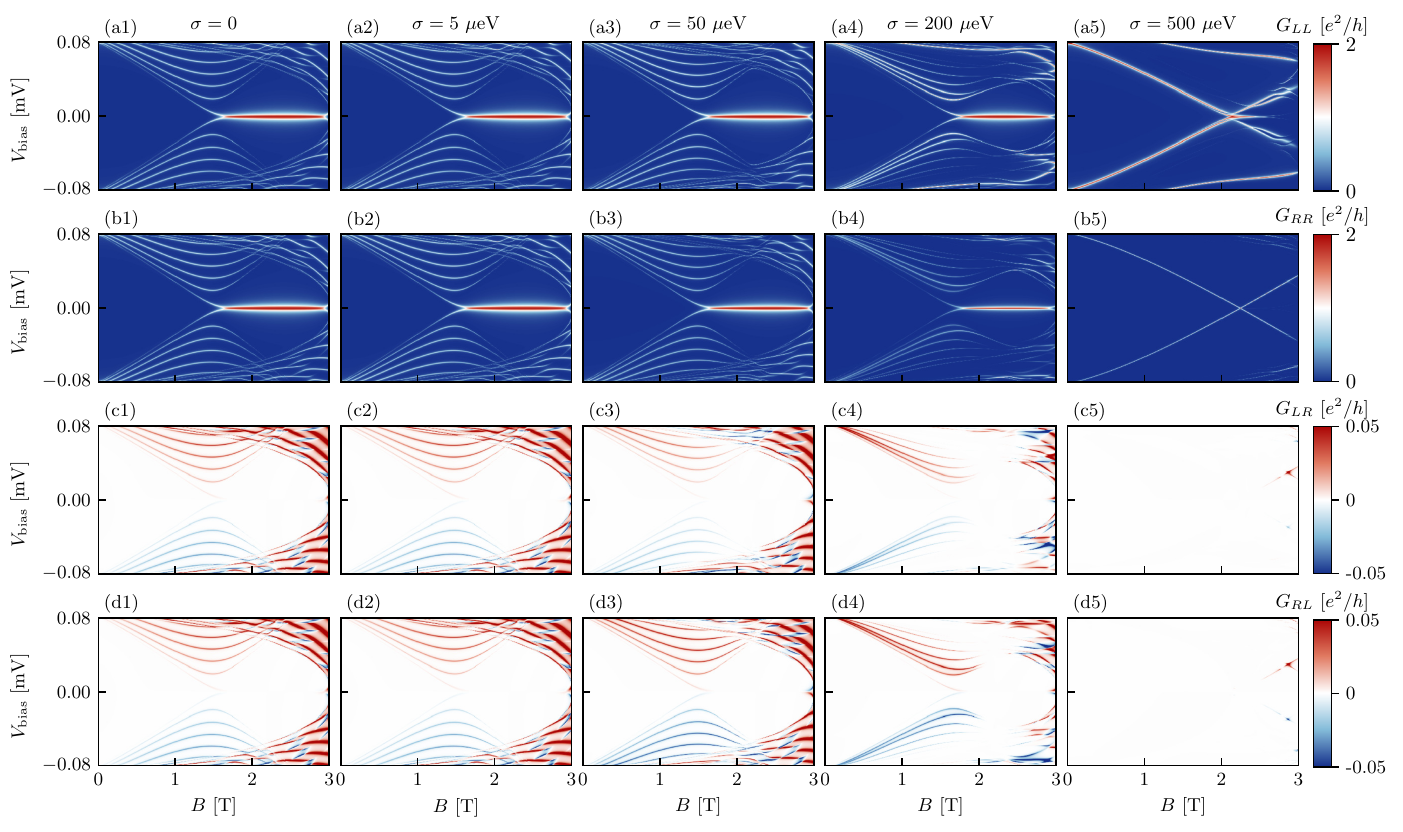}
	\caption{Local (first and second row) and nonlocal (third and fourth row) tunneling conductance spectra for a Ge/Al hybrid nanowire of length $L_x=3~\mu$m in the presence of random Gaussian disorder [see Eq.~(\ref{eq:disorder_uncorr})] for different disorder strengths $\sigma$. First column: $\sigma=0$, second column: $\sigma=5~\mu$eV, third column: $\sigma=50~\mu$eV, fourth column: $\sigma=200~\mu$eV, fifth column: $\sigma=500~\mu$eV. The Ge hole nanowire is modeled by Eq.~(\ref{eq:2band_sc}) using effective parameters extracted from the full multiband Hamiltonian (see Sec.~\ref{sec:model_ge_full}) with  $L_y=15$~nm, $L_z=22$~nm, $\mathcal{E}=0.5$~V$/\mu$m, $E_s=0.01$~eV, and $\tilde\mu_0=0$. The proximity-induced superconducting gap at zero magnetic field is set to $\Delta(0)=80~\mu$eV, and the critical field is taken to be $B_c=3$~T.}
	\label{fig:conductance_without_SE_L_3}
\end{figure*}

In Fig.~\ref{fig:conductance_without_SE_L_1} (Fig.~\ref{fig:conductance_without_SE_L_3}), we present results for a Ge/Al hybrid nanowire of length $L_x=1.5~\mu$m ($L_x=3~\mu$m) modeled by Eq.~(\ref{eq:2band_sc}), i.e., without the self-energy term. In this case, the topological gap is overestimated by a factor of $2-3$ compared to the case where self-energy effects are taken into account (see Figs.~\ref{fig:conductance_onsite_L_1} and \ref{fig:conductance_onsite_L_3} in the main text). In Figs.~\ref{fig:conductance_without_SE_L_1}(a1-d1) and \ref{fig:conductance_without_SE_L_3}(a1-d1), we show the differential conductance spectra in the pristine case. For both wire lengths, the left and right local conductance spectra manifest clear MZM-induced ZBPs beyond the topological phase transition, which occurs at a critical field $B^*\sim1.4$~T determined by the condition $V_{Z}^2(B^*)=\Delta^2(B^*)+\mu^2$. Increasing the disorder strength, we again find that the MZM-induced ZBPs in the local conductance as well as the closing and reopening of the bulk gap in the nonlocal conductance remain intact up to a disorder strength of about $\sim 200~\mu$eV, see Figs.~\ref{fig:conductance_without_SE_L_1}(a2-d2,a3-d3,a4-d4) and \ref{fig:conductance_without_SE_L_3}(a2-d2,a3-d3,a4-d4). Finally, at a disorder strength of $500~\mu$eV, the topology of the system becomes suppressed due to disorder. The conductance spectra now generally depend strongly on the specific disorder configuration, with some samples still manifesting features vaguely reminiscent of ZBPs [see Figs.~\ref{fig:conductance_without_SE_L_1}(a5,b5)] and some not [see Figs.~\ref{fig:conductance_without_SE_L_3}(a5,b5)]. For the shorter wire, the nonlocal conductances shown in Figs.~\ref{fig:conductance_without_SE_L_1}(c5,d5) show an apparent bulk gap closing, but no reopening. For the longer wire, the nonlocal conductance is now so strongly suppressed that Figs.~\ref{fig:conductance_without_SE_L_3}(c5,d5) appear featureless for our choice of color scale.

\section{Disorder-averaged conductance spectra}
\label{app:average}

\begin{figure*}[bt]
	\centering
	\includegraphics[width=0.7\textwidth]{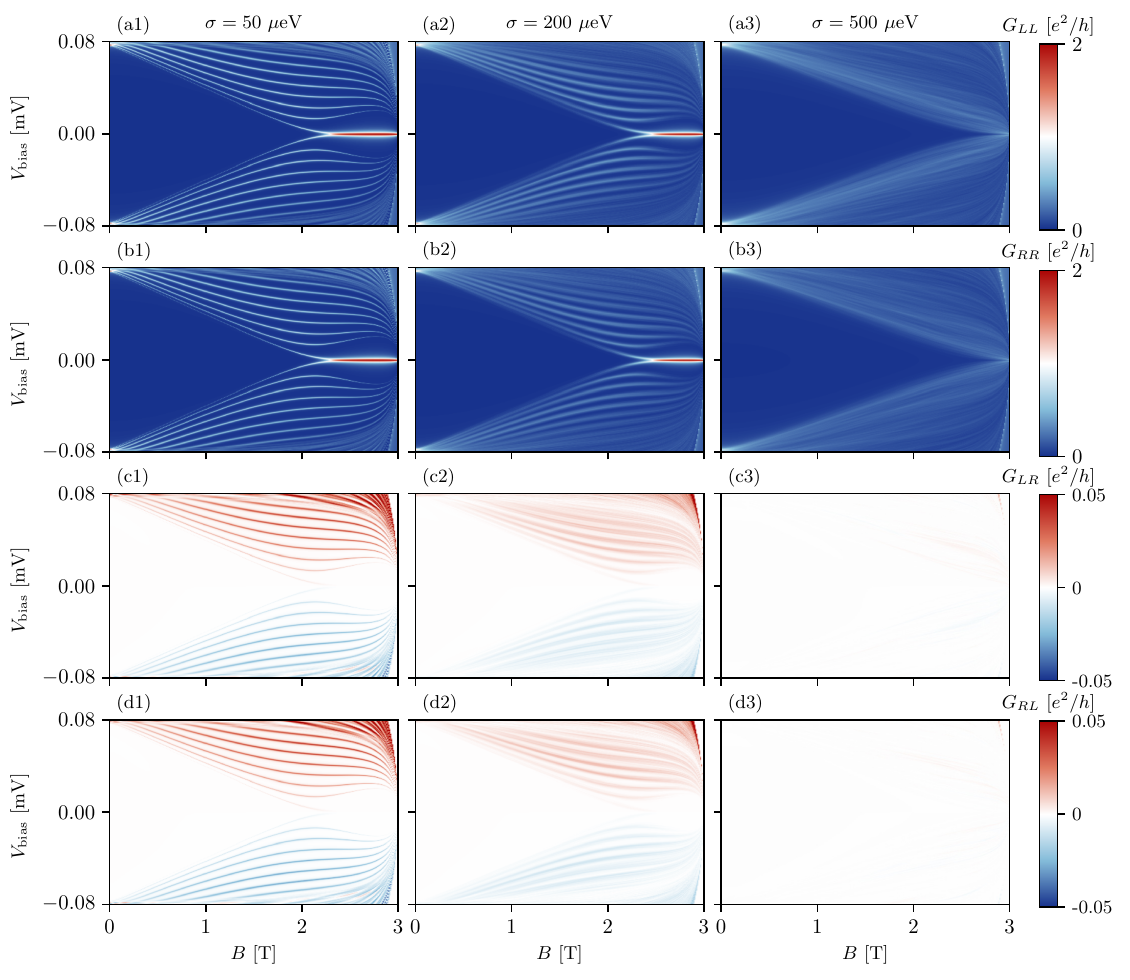}
	\caption{Disorder-averaged local (first and second row) and nonlocal (third and fourth row) tunneling conductance spectra for a Ge/Al hybrid nanowire of length $L_x=3~\mu$m in the presence of random Gaussian disorder [see Eq.~(\ref{eq:disorder_uncorr})]. First column: $\sigma=50~\mu$eV, second column: $\sigma=200~\mu$eV, third column: $\sigma=500~\mu$eV. The Ge hole nanowire is modeled by Eq.~(\ref{eq:2band_self}) using effective parameters extracted from the full multiband Hamiltonian (see Sec.~\ref{sec:model_ge_full}) with  $L_y=15$~nm, $L_z=22$~nm, $\mathcal{E}=0.5$~V$/\mu$m, $E_s=0.01$~eV, and $\tilde\mu_0=0$. Furthermore, we use $\Delta_0(0)=0.3$~meV for the superconducting gap of Al at zero magnetic field, $B_c=3$~T for the critical field, and $\gamma=0.1$~meV for the SC-SM coupling. We have averaged over 150 disorder configurations.}
	\label{fig:conductance_avg}
\end{figure*}

In Fig.~\ref{fig:conductance_avg}, we show the disorder-averaged tunneling conductance spectra for a Ge/Al hybrid nanowire of length $L_x=3~\mu$m in the presence of random Gaussian onsite disorder of strength $\sigma=50~\mu$eV [Figs.~\ref{fig:conductance_avg}(a1-d1)], $\sigma=200~\mu$eV [Figs.~\ref{fig:conductance_avg}(a2-d2)], and $\sigma=500~\mu$eV [Figs.~\ref{fig:conductance_avg}(a3-d3)]. In all panels, we have averaged over $150$ disorder configurations. As already mentioned in the main text, we find that, for $\sigma=50~\mu$eV, the disorder-averaged result looks almost indistinguishable from the pristine case. Furthermore, even for $\sigma=200~\mu$eV, the disorder-averaged conductance maps still exhibit all of the main qualitative features of the pristine case. In particular, the local tunneling conductance spectra  [see Figs.~\ref{fig:conductance_avg}(a2,b2)] still manifest well-discernible ZBPs (even though they generally emerge at slightly larger magnetic fields compared to the pristine case). Furthermore, the nonlocal tunneling conductance spectra  [see Figs.~\ref{fig:conductance_avg}(c2,d2)] still show a closing and reopening of the bulk gap, although the signal is visibly reduced compared to the pristine case. The disorder also generally reduces the size of the topological gap. Finally, for $\sigma=500~\mu$eV, the local conductance spectra [see Figs.~\ref{fig:conductance_avg}(a3,b3)] depend strongly on the specific disorder configuration and generally do not manifest any ZBPs. The nonlocal conductance signal is so strongly suppressed that Figs.~\ref{fig:conductance_avg}(c3,d3) appear featureless for our choice of color scale.

\section{Conductance spectra for varying impurity densities}
\label{app:ni}

\begin{figure*}[bt]
	\centering
	\includegraphics[width=1\textwidth]{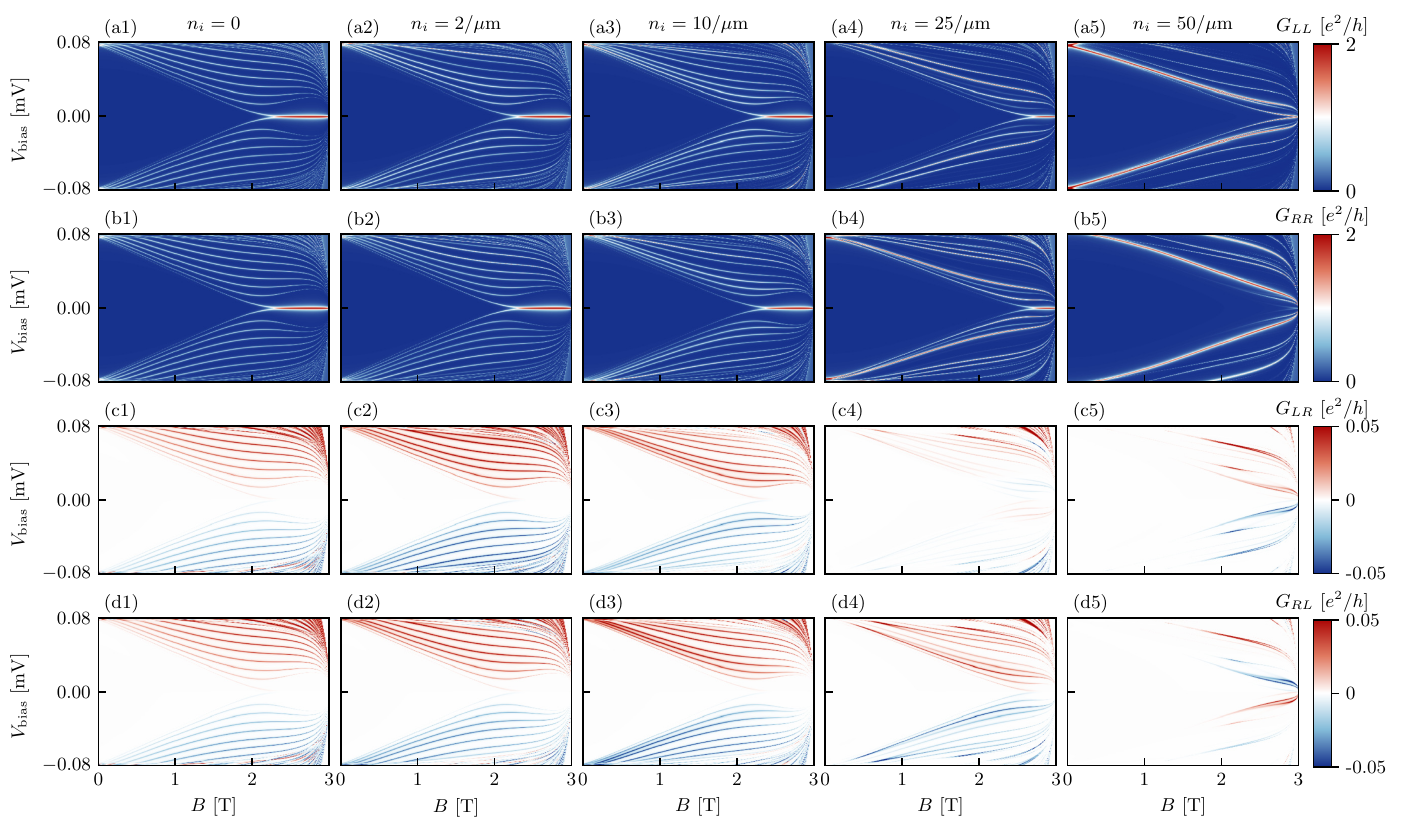}
	\caption{Local (first and second row) and nonlocal (third and fourth row) tunneling conductance spectra for a Ge/Al hybrid nanowire of length $L_x=3~\mu$m in the presence of correlated disorder [see Eq.~(\ref{eq:disorder_corr})] for different impurity densities $n_i$. First column: $n_i=0$, second column: $n_i=2/\mu$m, third column: $n_i=10/\mu$m, fourth column: $n_i=25/\mu$m, fifth column: $n_i=50/\mu$m. The Ge hole nanowire is modeled by Eq.~(\ref{eq:2band_self}) using effective parameters extracted from the full multiband Hamiltonian (see Sec.~\ref{sec:model_ge_full}) with $L_y=15$~nm, $L_z=22$~nm, $\mathcal{E}=0.5$~V$/\mu$m, $E_s=0.01$~eV, and $\tilde\mu_0=0$. Furthermore, we use $\Delta_0(0)=0.3$~meV for the superconducting gap of Al at zero magnetic field, $B_c=3$~T for the critical field, and $\gamma=0.1$~meV for the SC-SM coupling. For the disorder potential, we use the correlation length $\lambda=11$~nm and the impurity amplitude $V_0=200~\mu$eV.}
	\label{fig:conductance_correlated_ni}
\end{figure*}

In Fig.~\ref{fig:conductance_correlated_ni}, we present local and nonlocal tunneling conductance maps for a Ge/Al hybrid nanowire of length $L_x=3~\mu$m in the presence of spatially correlated disorder as described by Eq.~(\ref{eq:disorder_corr}). Here, we fix the correlation length to $\lambda= 11$~nm and the impurity amplitude to $V_0=200~\mu$eV, while we vary the impurity density $n_i$. We find that the topological phase survives up to an impurity density of $n_i\sim 25/\mu$m.

\section{Alternative versions of Figs.~\ref{fig:conductance_onsite_SE_NbTiN_L_1}(c5), \ref{fig:conductance_onsite_SE_NbTiN_L_1}(d5) and \ref{fig:conductance_onsite_SE_NbTiN_L_3}(c5), \ref{fig:conductance_onsite_SE_NbTiN_L_3}(d5)}

\begin{figure}[bt]
	\centering
	\includegraphics[width=1\columnwidth]{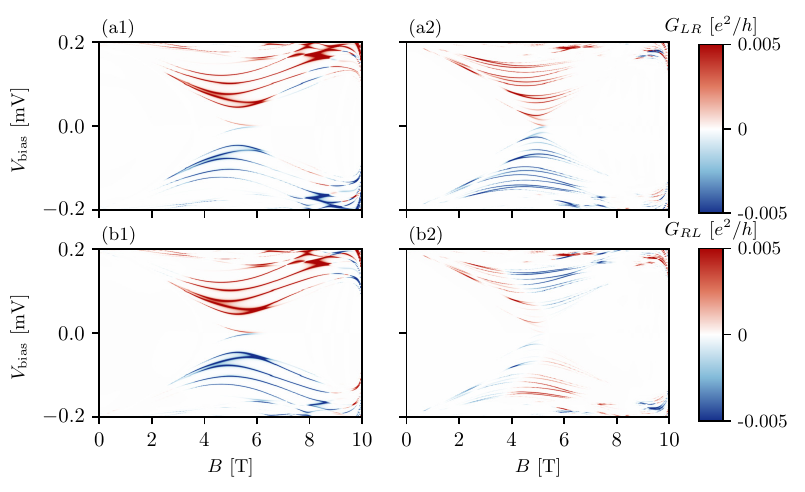}
	\caption{(a1,b1) and (a2,b2) show the same plots as Figs.~\ref{fig:conductance_onsite_SE_NbTiN_L_1}(c5,d5) and \ref{fig:conductance_onsite_SE_NbTiN_L_3}(c5,d5), respectively, but with a color bar that only ranges from $-0.005\, e^2/h$ to $0.005\, e^2/h$. In all panels, out-of-bound values are clipped.}
	\label{fig:conductance_NbTiN_zoom}
\end{figure}

In Fig.~\ref{fig:conductance_NbTiN_zoom}, we replot the nonlocal tunneling conductance maps shown in Figs.~\ref{fig:conductance_onsite_SE_NbTiN_L_1}(c5), \ref{fig:conductance_onsite_SE_NbTiN_L_1}(d5) and \ref{fig:conductance_onsite_SE_NbTiN_L_3}(c5), \ref{fig:conductance_onsite_SE_NbTiN_L_3}(d5) of the main text using a color scale that only ranges from $-0.005\, e^2/h$ to $0.005\, e^2/h$ (out-of-bound values clipped). We find that a gap closing and reopening can be observed if the precision of the nonlocal conductance measurement is high enough.

\begin{thebibliography}{2}%
\makeatletter
\providecommand \@ifxundefined [1]{%
 \@ifx{#1\undefined}
}%
\providecommand \@ifnum [1]{%
 \ifnum #1\expandafter \@firstoftwo
 \else \expandafter \@secondoftwo
 \fi
}%
\providecommand \@ifx [1]{%
 \ifx #1\expandafter \@firstoftwo
 \else \expandafter \@secondoftwo
 \fi
}%
\providecommand \natexlab [1]{#1}%
\providecommand \enquote  [1]{``#1''}%
\providecommand \bibnamefont  [1]{#1}%
\providecommand \bibfnamefont [1]{#1}%
\providecommand \citenamefont [1]{#1}%
\providecommand \href@noop [0]{\@secondoftwo}%
\providecommand \href [0]{\begingroup \@sanitize@url \@href}%
\providecommand \@href[1]{\@@startlink{#1}\@@href}%
\providecommand \@@href[1]{\endgroup#1\@@endlink}%
\providecommand \@sanitize@url [0]{\catcode `\\12\catcode `\$12\catcode
  `\&12\catcode `\#12\catcode `\^12\catcode `\_12\catcode `\%12\relax}%
\providecommand \@@startlink[1]{}%
\providecommand \@@endlink[0]{}%
\providecommand \url  [0]{\begingroup\@sanitize@url \@url }%
\providecommand \@url [1]{\endgroup\@href {#1}{\urlprefix }}%
\providecommand \urlprefix  [0]{URL }%
\providecommand \Eprint [0]{\href }%
\providecommand \doibase [0]{https://doi.org/}%
\providecommand \selectlanguage [0]{\@gobble}%
\providecommand \bibinfo  [0]{\@secondoftwo}%
\providecommand \bibfield  [0]{\@secondoftwo}%
\providecommand \translation [1]{[#1]}%
\providecommand \BibitemOpen [0]{}%
\providecommand \bibitemStop [0]{}%
\providecommand \bibitemNoStop [0]{.\EOS\space}%
\providecommand \EOS [0]{\spacefactor3000\relax}%
\providecommand \BibitemShut  [1]{\csname bibitem#1\endcsname}%
\let\auto@bib@innerbib\@empty
%</preamble>
\bibitem [{Note1()}]{Note1}%
  \BibitemOpen
  \bibinfo {note} {The strain energy depends approximately linearly on the
  percentage of Si in the SiGe barrier. Reference~\cite {Sammak2019} reports
  $\varepsilon _{||}=-0.63\%$ in a Ge/SiGe quantum well with $20\%$ of Si in
  the barrier, from which we estimate $E_s=-|b|\varepsilon _0\approx 0.0237$~eV
  using $b\approx -2.16$~eV~\cite {Bir1974} and $\varepsilon _0\approx
  1.74\varepsilon _{||}$~\cite {Terrazos2021}. Extrapolating the linear
  dependence, our choice of $E_s\approx 0.01$~eV corresponds to $\approx $
  8.5\% of Si in the barrier, which is rather optimistic but not unrealistic
  compared to current state-of-the-art devices, where the Si content is
  typically 10\%-20\%.}\BibitemShut {Stop}%
\bibitem [{Note2()}]{Note2}%
  \BibitemOpen
  \bibinfo {note} {We note that the nonlinearity in $\omega $ appearing in the
  strong coupling limit does not further complicate the numerical procedure
  since we are evaluating the transmission coefficients for a discrete set of
  energies, such that, for each fixed energy, the Hamiltonian is a purely
  numerical matrix.}\BibitemShut {Stop}%
\end{thebibliography}%


\begin{thebibliography}{}
		
		
	%disorder problem in conventional semiconductor nanowires
	\bibitem{Dassarma2023}
	S. Das Sarma, In search of Majorana, Nat. Phys. {\bf 19}, 165 (2023).

	
	%conventional Majorana nanowire
	\bibitem{Lutchyn2010}
	R. M. Lutchyn, J. D. Sau, and S. Das Sarma, Majorana Fermions and a Topological Phase Transition in Semiconductor-Superconductor Heterostructures, Phys. Rev. Lett. {\bf 105}, 077001 (2010).
	\bibitem{Oreg2010}
	Y. Oreg, G. Refael, and F. von Oppen, Helical Liquids and Majorana Bound States in Quantum Wires, Phys. Rev. Lett. {\bf 105}, 177002 (2010).
	\bibitem{Stanescu2011}
	T. D. Stanescu, R. M. Lutchyn, and S. Das Sarma, Majorana fermions in semiconductor nanowires, Phys. Rev. B {\bf 84}, 144522 (2011).
	
	%Majorana experiment
	\bibitem{Mourik2012}
	V. Mourik, K. Zuo, S. M. Frolov, S. R. Plissard, E. P. A. M. Bakkers, and L. P. Kouwenhoven, Signatures of Majorana fermions in hybrid superconductor-semiconductor nanowire devices, Science {\bf 336}, 6084 (2012).
	\bibitem{Rohkinson2012}
	L. P. Rokhinson, X. Liu, and J. K. Furdyna, The fractional a.c. Josephson effect in a semiconductor–superconductor nanowire as a signature of Majorana particles, Nat. Phys. {\bf 8}, 795 (2012).
	\bibitem{Das2012}
	A. Das, Y. Ronen, Y. Most, Y. Oreg, M. Heiblum, and H. Shtrikman, Zero-bias peaks and splitting in an Al–InAs nanowire topological superconductor as a signature of Majorana fermions, Nat. Phys. {\bf 8}, 887 (2012).
	\bibitem{Deng2012}
	M. T. Deng, C. L. Yu, G. Y. Huang, M. Larsson, P. Caroff, and H. Q. Xu, Anomalous Zero-Bias Conductance Peak in a Nb–InSb Nanowire–Nb Hybrid Device, Nano Lett. {\bf 12}, 6414 (2012).
	\bibitem{Lee2012}
	E. J. H. Lee, X. Jiang, R. Aguado, G. Katsaros, C. M. Lieber, and S. De Franceschi, Zero-Bias Anomaly in a Nanowire Quantum Dot Coupled to Superconductors, Phys. Rev. Lett. {\bf 109}, 186802 (2012).
	\bibitem{Churchill2013}
	H. O. H. Churchill, V. Fatemi, K. Grove-Rasmussen, M. T. Deng, P. Caroff, H. Q. Xu, and C. M. Marcus, Superconductor-nanowire devices from tunneling to the multichannel regime: Zero-bias oscillations and magnetoconductance crossover, Phys. Rev. B {\bf 87}, 241401(R) (2013).
	\bibitem{Deng2016}
	M. T. Deng, S. Vaitiekenas, E. B. Hansen, J. Danon, M. Leijnse, K. Flensberg, J. Nyg\r{a}rd, P. Krogstrup, and C. M. Marcus, Majorana bound state in a coupled quantum-dot hybrid-nanowire system, Science {\bf 354}, 1557 (2016).
	\bibitem{Deng2018}
	M. T. Deng, S. Vaitiekenas, E. Prada, P. San-Jose, J. Nyg\r{a}rd, P. Krogstrup, R. Aguado, and C. M. Marcus, Nonlocality of Majorana modes in hybrid nanowires, Phys. Rev. B {\bf 98}, 085125 (2018).
	
	\bibitem{Sau2012}
	J. D. Sau, S. Tewari, and S. Das Sarma, Experimental and materials considerations for the topological superconducting state in electron- and hole-doped semiconductors: Searching for non-Abelian Majorana modes in 1D nanowires and 2D heterostructures, Phys. Rev. B {\bf 85}, 064512 (2012).
	\bibitem{Sau2013}
	J. D. Sau and S. Das Sarma, Density of states of disordered topological superconductor-semiconductor hybrid nanowires, Phys. Rev. B {\bf 88}, 064506 (2013).
	\bibitem{Chiu2017}
	C.-K. Chiu, J. D. Sau, and S. Das Sarma, Conductance of a superconducting Coulomb-blockaded Majorana nanowire, Phys. Rev. B {\bf 96}, 054504 (2017).
	\bibitem{Pan2020}
	H. Pan and S. Das Sarma, Physical mechanisms for zero-bias conductance peaks in Majorana nanowires, Phys. Rev. Research {\bf 2}, 013377 (2020).
	\bibitem{Pan2020b}
	H. Pan, W. S. Cole, J. D. Sau, and S. Das Sarma, Generic quantized zero-bias conductance peaks in superconductor-semiconductor hybrid structures, Phys. Rev. B {\bf 101}, 024506 (2020).
	\bibitem{Ahn2021}
	S. Ahn, H. Pan, B. Woods, T. D. Stanescu, and S. Das Sarma, Estimating disorder and its adverse effects in semiconductor Majorana nanowires, Phys. Rev. Materials {\bf 5}, 124602 (2021).
	\bibitem{Dassarma2021}
	S. Das Sarma and H. Pan, Disorder-induced zero-bias peaks in Majorana nanowires, Phys. Rev. B {\bf 103}, 195158 (2021).
	\bibitem{Pan2021}
	H. Pan, J. D. Sau, and S. Das Sarma, Three-terminal nonlocal conductance in Majorana nanowires: Distinguishing topological and trivial in realistic systems with disorder and inhomogeneous potential, Phys. Rev. B {\bf 103}, 014513 (2021).
	
	%MSFT experiment
	\bibitem{Aghaee2022}
	M. Aghaee \emph{et al.}, InAs-Al hybrid devices passing the topological gap protocol, Phys. Rev. B {\bf 107}, 245423 (2023).

	%theory for MSFT
	\bibitem{Dassarma2023b}
	S. Das Sarma, J. D. Sau, and T. D. Stanescu, Spectral properties, topological patches, and effective phase diagrams of finite disordered Majorana nanowires, Phys. Rev. B {\bf 108}, 085416 (2023).
	\bibitem{Dassarma2023c}
	S. Das Sarma and H. Pan, Density of states, transport, and topology in disordered Majorana nanowires, Phys. Rev. B {\bf 108}, 085415 (2023).
	
	%review on Ge
	\bibitem{Scappucci2021}
	G. Scappucci, C. Kloeffel, F. A. Zwanenburg, D. Loss, M. Myronov, J.-J. Zhang, S. De Franceschi, G. Katsaros, and M. Veldhorst, The germanium quantum information route, Nat. Rev. Mater. {\bf 6}, 926 (2021).
	
	
	%high-quality Ge
	\bibitem{Sammak2019}
	A. Sammak, D. Sabbagh, N. W. Hendrickx, M. Lodari, B. P. Wuetz, A. Tosato, L. Yeoh, M. Bollani, M. Virgilio, M. A. Schubert, P. Zaumseil, G. Capellini, M. Veldhorst, and G. Scappucci, Shallow and Undoped Germanium Quantum Wells: A Playground for Spin and Hybrid Quantum Technology, Adv. Funct. Mater. {\bf 29}, 1807613 (2019).
	\bibitem{Lodari2022}
	M. Lodari, O. Kong, M. Rendell, A. Tosato, A. Sammak, M. Veldhorst, A. R. Hamilton, and G. Scappucci, Lightly strained germanium quantum wells with hole mobility exceeding one million, Appl. Phys. Lett. {\bf 120}, 122104 (2022).
	\bibitem{Myronov2023}
	M. Myronov, J. Kycia, P. Waldron, W. Jiang, P. Barrios, A. Bogan, P. Coleridge, and S. Studenikin, Holes Outperform Electrons in Group IV Semiconductor Materials, Small
	Science {\bf 3}, 2200094 (2023).
	\bibitem{Stehouwer2023}
	L. E. A. Stehouwer, A. Tosato, D. Degli Esposti, D. Costa, M. Veldhorst, A. Sammak, G. Scappucci, Germanium wafers for strained quantum wells with low disorder, Appl. Phys. Lett. {\bf 123}, 092101 (2023).
	
	
	%proximity-induced SC in Germanium
	\bibitem{Vries2018}
	F. K. de Vries, J. Shen, R. J. Skolasinski, M. P. Nowak, D. Varjas, L. Wang, M. Wimmer, J. Ridderbos, F. A. Zwanenburg, A. Li, S. Koelling, M. A. Verheijen, E. P. A. M. Bakkers, and L. P. Kouwenhoven, Spin–Orbit Interaction and Induced Superconductivity in a One-Dimensional Hole Gas, Nano Lett. {\bf 18}, 6483 (2018).
	\bibitem{Hendrickx2018}
	N. Hendrickx, D. Franke, A. Sammak, M. Kouwenhoven, D. Sabbagh, L. Yeoh, R. Li, M. Tagliaferri, M. Virgilio, G. Capellini, G. Scappucci, and M. Veldhorst, Gate-controlled quantum dots and superconductivity in planar germanium, Nat. Comm. {\bf 9}, 2835 (2018).
	\bibitem{Hendrickx2019}
	N. W. Hendrickx, M. L. V. Tagliaferri, M. Kouwenhoven, R. Li, D. P. Franke, A. Sammak, A. Brinkman, G. Scappucci, and M. Veldhorst, Ballistic supercurrent discretization and micrometer-long Josephson coupling in germanium, Phys. Rev. B {\bf 99}, 075435 (2019).
	\bibitem{Vigneau2019}
	F. Vigneau, R. Mizokuchi, D. C. Zanuz, X. Huang, S. Tan, R. Maurand, S. Frolov, A. Sammak, G. Scappucci, F. Lefloch, and S. De Franceschi, Germanium Quantum-Well Josephson Field-Effect Transistors and Interferometers, Nano Lett. {\bf 19}, 1023 (2019).
	\bibitem{Valentini2024}
	M. Valentini, O. Sagi, L. Baghumyan, T. de Gijsel, J. Jung, S. Calcaterra, A. Ballabio, J. Aguilera Servin, K. Aggarwal, M. Janik, T. Adletzberger, R. Seoane Souto, M. Leijnse, J. Danon, C. Schrade, E. Bakkers, D. Chrastina, G. Isella, and G. Katsaros, Parity-conserving Cooper-pair transport and ideal superconducting diode in planar germanium, Nature Comm. {\bf 15}, 169 (2024).
	\bibitem{Aggarwal2021}
	K. Aggarwal, A. Hofmann, D. Jirovec, I. Prieto, A. Sammak, M. Botifoll, S. Mart\'{i}-S\'{a}nchez, M. Veldhorst, J. Arbiol, G. Scappucci, J. Danon, and G. Katsaros, Enhancement of proximity-induced superconductivity in a planar Ge hole gas, Phys. Rev. Res. {\bf 3}, L022005 (2021).
	\bibitem{Tosato2022}
	A. Tosato, V. Levajac, J.-Y. Wang, C. J. Boor, F. Borsoi, M. Botifoll, C. N. Borja, S. Mart\'{i}-S\'{a}nchez, J. Arbiol, A. Sammak, M. Veldhorst, and G. Scappucci, Hard superconducting gap in germanium, Commun. Mater. {\bf 4}, 23 (2023).
	

	%Majoranas in Ge hole channels
	\bibitem{Laubscher2023}
	K. Laubscher, J. D. Sau, and S. Das Sarma, Majorana zero modes in gate-defined germanium hole nanowires, Phys. Rev. B {\bf 109}, 035433 (2024).
	
	
	%Basic properties of Ge
	\bibitem{Luttinger1956}
	J. M. Luttinger, Phys. Rev. {\bf 102}, Quantum Theory of Cyclotron Resonance in Semiconductors: General Theory, 1030 (1956).
	\bibitem{Winkler2003}
	R. Winkler, {\it Spin-orbit coupling effects in two-dimensional electron and hole systems} (Springer-Verlag, Berlin, Heidelberg, New York, 2003).
	
	
	%strain
	\bibitem{Bir1974}
	G. L. Bir, G. E. Pikus, \emph{et al.}, {\it Symmetry and strain-induced effects in semiconductors}, Vol. 484 (Wiley, New
	York, 1974).
	
	
	%kappa for Ge
	\bibitem{Lawaetz1971}
	P. Lawaetz, Valence-Band Parameters in Cubic Semiconductors, Phys. Rev. B {\bf 4}, 3460 (1971).
	
	\bibitem{Adelsberger2022}
	C. Adelsberger, S. Bosco, J. Klinovaja, and D. Loss, Enhanced orbital magnetic field effects in Ge hole nanowires, Phys. Rev. B {\bf 106}, 235408 (2022).	
	
	%Ge hole wires: theory
	\bibitem{Csontos2009}%HH-LH mixing in hole nanowires
	D. Csontos, P. Brusheim, U. Z\"{u}licke, and H. Q. Xu, Spin-3/2 physics of semiconductor hole nanowires: Valence-band mixing and tunable interplay between bulk-material and orbital bound-state spin splittings, Phys. Rev. B {\bf 79}, 155323 (2009).
	\bibitem{Kloeffel2011}%direct SOI
	C. Kloeffel, M. Trif, and D. Loss, Strong spin-orbit interaction and helical hole states in Ge/Si nanowires, Phys. Rev. B {\bf 84}, 195314 (2011).
	\bibitem{Kloeffel2018}%direct SOI
	C. Kloeffel, M. J. Ran\v{c}i\'{c}, and D. Loss, Direct Rashba spin-orbit interaction in Si and Ge nanowires with different growth directions, Phys. Rev. B {\bf 97}, 235422 (2018).
	\bibitem{Milivojevic2021}
	M. Milivojevi\'{c}, Electrical control of the hole spin qubit in Si and Ge nanowire quantum dots, Phys. Rev. B {\bf 104}, 235304 (2021).

	\bibitem{Adelsberger2021}
	C. Adelsberger, M. Benito, S. Bosco, J. Klinovaja, and D. Loss, Hole-spin qubits in Ge nanowire quantum dots: Interplay of orbital magnetic field, strain, and growth direction, Phys. Rev. B {\bf 105}, 075308 (2022).
	
	\bibitem{Terrazos2021}
	L. A. Terrazos, E. Marcellina, Z. Wang, S. N. Coppersmith, M. Friesen, A. R. Hamilton, X. Hu, B. Koiller, A. L. Saraiva, D. Culcer, and R. B. Capaz, Theory of hole-spin qubits in strained germanium quantum dots, Phys. Rev. B {\bf 103}, 125201 (2021).
	
	%proximity-induced SC in hole nanowire: equal pairing amplitudes
	\bibitem{Mao2012}
	L. Mao, M. Gong, E. Dumitrescu, S. Tewari, and C. Zhang, Hole-Doped Semiconductor Nanowire on Top of an s-Wave Superconductor: A New and Experimentally Accessible System for Majorana Fermions, Phys. Rev. Lett. {\bf 108}, 177001 (2012).
	
	
	%proximity-induced SC: tunneling between SC and material
	\bibitem{Sau2010}
	J. D. Sau, R. M. Lutchyn, S. Tewari, and S. Das Sarma, Robustness of Majorana fermions in proximity-induced superconductors, Phys. Rev. B {\bf 82}, 094522 (2010).
	\bibitem{Stanescu2010}
	T. D. Stanescu, J. D. Sau, R. M. Lutchyn, and S. Das Sarma, Proximity effect at the superconductor–topological insulator interface, Phys. Rev. B {\bf 81}, 241310(R) (2010).
	\bibitem{Cole2015}
	W. S. Cole, S. Das Sarma, and T. D. Stanescu, Effects of large induced superconducting gap on semiconductor Majorana nanowires, Phys. Rev. B {\bf 92}, 174511 (2015).
	\bibitem{Reeg2018}
	C. Reeg, D. Loss, and J. Klinovaja, Metallization of a Rashba wire by a superconducting layer in the strong-proximity regime, Phys. Rev. B {\bf 97}, 165425 (2018).
	\bibitem{Stanescu2013}
	T. D. Stanescu and S. Tewari, Majorana fermions in semiconductor nanowires, J. Phys.: Condens. Matter {\bf 25}, 233201 (2013).
	
	
	% proximity effect for holes
	\bibitem{Adelsberger2023}
	C. Adelsberger, H. F. Legg, D. Loss, and J. Klinovaja, Microscopic analysis of proximity-induced superconductivity and metallization effects in superconductor-germanium hole nanowires, Phys. Rev. B {\bf 108}, 155433 (2023).
	\bibitem{Moghaddam2014}
	A. G. Moghaddam, T. Kernreiter, M. Governale, and U. Z\"{u}licke, Exporting superconductivity across the gap: Proximity effect for semiconductor valence-band states due to contact with a simple-metal superconductor, Phys. Rev. B {\bf 89}, 184507 (2014).
	
	
	%critical field of Al
	\bibitem{Nichele2017}
	F. Nichele, A. C. C. Drachmann, A. M. Whiticar, E. C. T. O'Farrell, H. J. Suominen, A. Fornieri, T. Wang, G. C. Gardner, C. Thomas, A. T. Hatke, P. Krogstrup, M. J. Manfra, K. Flensberg, and C. M. Marcus, Scaling of Majorana Zero-Bias Conductance Peaks, Phys. Rev. Lett. {\bf 119}, 136803 (2017).
	\bibitem{Suominen2017}
	H. J. Suominen, M. Kjaergaard, A. R. Hamilton, J. Shabani, C. J. Palmstr\o{}m, C. M. Marcus, and F. Nichele, Zero-Energy Modes from Coalescing Andreev States in a Two-Dimensional Semiconductor-Superconductor Hybrid Platform, Phys. Rev. Lett. {\bf 119}, 176805 (2017).
	
	
	%bulk gaps of Al an NbTiN
	\bibitem{Lutchyn2018}
	R. M. Lutchyn, E. P. A. M. Bakkers, L. P. Kouwenhoven, P. Krogstrup, C. M. Marcus, and Y. Oreg, Majorana zero modes in superconductor–semiconductor heterostructures, Nature Rev. Mater. {\bf 3}, 52 (2018).
	
	
	%charge impurities in MZM nanowires
	\bibitem{Woods2021}
	B. D. Woods, S. Das Sarma, and T. D. Stanescu, Charge-Impurity Effects in Hybrid Majorana Nanowires, Phys. Rev. Applied {\bf 16}, 054053 (2021).
	
	
	%tunnel barrier
	\bibitem{Setiawan2017}
	F. Setiawan, C.-X. Liu, J. D. Sau, and S. Das Sarma, Electron temperature and tunnel coupling dependence of zero-bias and almost-zero-bias conductance peaks in Majorana nanowires, Phys. Rev. B {\bf 96}, 184520 (2017).
	\bibitem{Liu2017}
	C.-X. Liu, F. Setiawan, J. D. Sau, and S. Das Sarma, Phenomenology of the soft gap, zero-bias peak, and zero-mode splitting in ideal Majorana nanowires, Phys. Rev. B {\bf 96}, 054520 (2017).
	
	
	%dissipation
	\bibitem{Liu2017b}
	C.-X. Liu, J. D. Sau, and S. Das Sarma, Role of dissipation in realistic Majorana nanowires, Phys. Rev. B {\bf 95}, 054502 (2017).
	
	\bibitem{Blonder1982}
	G. E. Blonder, M. Tinkham, and T. M. Klapwijk, Transition from metallic to tunneling regimes in superconducting microconstrictions: Excess current, charge imbalance, and supercurrent conversion, Phys. Rev. B {\bf 25}, 4515 (1982).
	
	
	%KWANT
	\bibitem{Groth2014}
	C. W. Groth, M. Wimmer, A. R. Akhmerov, and X. Waintal, Kwant: a software package for quantum transport, New J. Phys. {\bf 16}, 063065 (2014).
	
	
	%nonlocal conductance
	\bibitem{Rosdahl2018}
	T. \"{O}. Rosdahl, A. Vuik, M. Kjaergaard, and A. R. Akhmerov, Andreev rectifier: A nonlocal conductance signature of topological phase transitions, Phys. Rev. B {\bf 97}, 045421 (2018).
	
	\bibitem{Menard2020}
	G. C. Menard, G. L. R. Anselmetti, E. A. Martinez, D. Puglia, F. K. Malinowski, J. S. Lee, S. Choi, M. Pendharkar, C. J. Palmstr\o{}m, K. Flensberg, C. M. Marcus, L. Casparis, and A. P. Higginbotham, Conductance-Matrix Symmetries of a Three-Terminal Hybrid Device, Phys. Rev. Lett. {\bf 124}, 036802 (2020).
	\bibitem{Puglia2021}
	D. Puglia, E. A. Martinez, G. C. M\'{e}nard, A. P\"{o}schl, S. Gronin, G. C. Gardner, R. Kallaher, M. J. Manfra, C. M. Marcus, A. P. Higginbotham, and L. Casparis, Closing of the induced gap in a hybrid superconductor-semiconductor nanowire, Phys. Rev. B {\bf 103}, 235201 (2021).
	
	
	%NbTiN on Si/Ge
	\bibitem{Su2016}
	Z. Su, A. Zarassi, B.-M. Nguyen, J. Yoo, S. A. Dayeh, and S. M. Frolov, High critical magnetic field superconducting contacts to Ge/Si core/shell nanowires, arXiv:1610.03010.
	
	
	%Ta as a good SC for Ge
	\bibitem{Strohbeen2023b}
	P. J. Strohbeen, A. M. Brook, I. Levy, W. L. Sarney, J. van Dijk, H. Orth, M. Mikalsen, and J. Shabani, Molecular beam epitaxy growth of superconducting tantalum germanide, Appl. Phys. Lett. {\bf 124}, 092102 (2024).
	%hyperdoped SiGe as a good SC for Ge
	\bibitem{Strohbeen2023}
	P. J. Strohbeen, A. M. Brook, W. L. Sarney, and J. Shabani, Superconductivity in hyperdoped Ge by molecular beam epitaxy, AIP Adv. {\bf 13}, 085118 (2023).
	
	
	%alternative ge-based mzm platforms
	\bibitem{Luethi2022}
	M. Luethi, K. Laubscher, S. Bosco, D. Loss, and J. Klinovaja, Planar Josephson junctions in germanium: Effect of cubic spin-orbit interaction, Phys. Rev. B {\bf 107}, 035435 (2023).
	\bibitem{Luethi2023}
	M. Luethi, H. F. Legg, K. Laubscher, D. Loss, and J. Klinovaja, Majorana bound states in germanium Josephson junctions via phase control, Phys. Rev. B {\bf 108}, 195406 (2023).
	
	
	%orbital effects in electron nanowires
	\bibitem{Dmytruk2018}
	O. Dmytruk and J. Klinovaja, Suppression of the overlap between Majorana fermions by orbital magnetic effects in semiconducting-superconducting nanowires, Phys. Rev. B {\bf 97}, 155409 (2018).
	
	
	
	
\end{thebibliography}
\end{document}